\documentclass{aa}
\usepackage{graphicx}
\usepackage{txfonts}
\begin{document}

\title{Mass-loss and diffusion in subdwarf B stars and \\
hot white dwarfs: do weak winds exist?}

   \author{K. Unglaub}
%          
%\fnmsep\thanks{Just to show the usage
%          of the elements in the author field}

   \offprints{K. Unglaub, \email{unglaub@sternwarte.uni-erlangen.de}}
 
   \institute{Dr. Remeis-Sternwarte, Astronomisches Institut der Universit\"at 
              Erlangen-N\"urnberg,
              Sternwartstrasse 7, D-96049 Bamberg, Germany
             \email{unglaub@sternwarte.uni-erlangen.de}
             }
 
   \date{}

\abstract
{According to previous investigations, the effect of diffusion in the stellar 
atmospheres and envelopes of hot white dwarfs and subdwarf B (sdB) stars 
strongly depends on the presence of weak winds with mass-loss rates  
$\dot M < 10^{-11} M_{\odot}/ \rm yr$.} 
{As in most of these stars with luminosities $L /L_{\odot} \la 100$, 
no wind signatures have been detected, the mass-loss rates are unknown.
In the present paper mass-loss rates are predicted from 
the original theory of radiatively driven winds.} 
{The method of solution is 
modified  so that the usual parametrization of the line force multipliers 
is not necessary. This is important especially for very thin winds.
In addition we checked whether a one-component description is 
justified. As a consequence of various simplifications, the mass-loss rates are 
expected to be overestimated.}
{Results are presented for effective temperatures in the range 
$25000 \, \rm K \leq T_{\rm eff} \leq 50000 \, \rm K$ and for various metallicities 
between solar and $Z / Z_{\odot} = 0.01$. For (pre-) white dwarfs and sdB 
stars a stellar mass of $M_{*} = 0.5 M_{\odot}$ is assumed. For fixed 
values of $T_{\rm eff}$, $M_{*}$, and $Z$, the results predict decreasing 
mass-loss rates with increasing surface gravity and an increasing 
dependence of the mass-loss rates on the metallicity. For white 
dwarfs with $\log g > 7.0$ no wind solution exists even if the metallicity 
would be solar. Winds with mass-loss rates around $10^{-11}$ to 
$10^{-10} M_{\odot} / \rm yr$ are predicted for the most luminous sdB stars 
with surface gravities of $\log g \la 5.5$, if the metallicity is not significantly 
lower than solar. For lower values of $\dot M$ 
metals decouple from hydrogen and helium.} 
{If weak winds with $\dot M \la 10^{-12} M_{\odot} / \rm yr$ exist, the metals 
cannot be coupled to hydrogen and helium. This should lead to additional changes 
in the surface composition, which have not yet been taken into account in the diffusion 
calculations with and without mass-loss. A possible scenario is the existence 
of pure metallic winds with mass-loss rates of $\dot M \la 10^{-16} M_{\odot} / \rm yr$
and with hydrostatic hydrogen and helium.}


 \titlerunning{Weak winds in sdB stars and hot white dwarfs}
 
 \authorrunning{Unglaub}
 
   \maketitle

%+++++++++++++++++++++++++++++++++++++++++++++++++++++++++++++++++
\section{Introduction}
%+++++++++++++++++++++++++++++++++++++++++++++++++++++++++++++++++
The abundance anomalies in several types of chemically peculiar stars are 
believed to be at least partially due to diffusion in the stellar 
atmosphere and envelope. The predictions of diffusion calculations strongly 
depend on the presence of mass-loss.
In several papers (Unglaub \& Bues \cite{ub98}, \cite{ub00}, \cite{ub01}), we 
investigated the combined effects of gravitational settling, radiative 
levitation, and weak winds on 
the chemical composition of hot white dwarfs and subdwarf B (sdB) stars. 
According to the results, these effects can explain the decreasing 
abundances of helium and metals during the cooling process of white dwarfs 
on the upper cooling sequence (with effective temperatures 
$T_{\rm eff} \ga 50000 \, \rm K$).
The mass-loss rates were estimated either from scaling laws or were 
a free parameter. 
The winds were assumed to be ``chemically homogeneous".
If $\dot M_{l}$ is the mass-loss rate of an element and $\zeta_{l}$ 
its mass fraction in the photosphere, then this assumption states that 
$\dot M_{l} = \zeta_{l} \dot M$, where $\dot M$ is the total mass-loss 
rate. Such a wind prevents (if $\dot M \ga 10^{-11} M_{\odot} / \rm yr$) or 
retards (if $\dot M < 10^{-11} M_{\odot} / \rm yr$) gravitational settling.
The present paper investigates whether the assumed mass-loss rates 
are consistent with the predictions of the theory of radiation-driven winds
and whether these winds may be chemically homogeneous.

In thin winds the one-component description is questionable, because 
the momentum redistribution via Coulomb 
collisions between metals (which are preferably accelerated due to the 
radiative flux) and hydrogen and helium may not be effective enough (e.g. 
Owocki \& Puls \cite{owo02}; Krti\v cka et al. \cite{krt03}). This may lead 
to selective winds in which the metals are expelled from the 
star, whereas hydrogen and helium are left behind (Babel \cite{bab95}, 
\cite{bab96}). In contrast to chemically homogeneous winds, selective winds 
should directly change the surface composition. 

In the absence of 
any mass-loss and other disturbing processes (e.g. convective mixing), 
an equilibrium between gravitational settling and radiative levitation should 
be expected. Because of saturation effects the radiative force on an element in 
the stellar atmosphere and envelope depends on its abundance. 
The radiative force decreases with increasing abundance, so an ``equilibrium abundance" can be found, 
for which the radiative force balances the effect of gravitational settling. 
This theory is described in Chayer et al. (\cite{chay9a}; \cite{chay9b}) and 
references therein. Dreizler \& Wolff (\cite{drei99}) and Schuh et al. (\cite{son02}) 
have incorporated the diffusion theory into their NLTE stellar atmosphere code,  
so synthetic spectra have been calculated for an abundance stratification, which 
is given from the equilibrium condition between gravitational 
settling and radiative levitation. 
However, for hot white dwarfs the quantitative agreement 
with observational results is not satisfactory in many cases. The presence of weak winds 
may be a possible reason for these discrepancies.
 
Winds have been detected in pre-white dwarfs with effective temperatures in 
the range $30000 \, \rm  K \la T_{\rm eff} \la 150000 \, \rm K$ and surface 
gravities $3.5 \la \log g \la 6.0$ (cgs units). These objects are in a 
post-asymptotic giant branch stage of evolution, and they evolve with 
approximately constant luminosities close to $10^{4} L_{\odot}$
to higher effective temperatures. As derived from observations and 
theoretical calculations (see e.g. Kudritzki et al. \cite{kud06}; 
Pauldrach et al. \cite{paul04};
Tinkler \& Lamers \cite{tin02}; Herald et al. \cite{her05}; 
Koesterke et al. \cite{koe98}; Koesterke \& Werner \cite{jkoe98} and the 
review of Kudritzki \& Puls \cite{kud00}), the mass-loss rates are between 
$10^{-6}$ and $10^{-9} M_{\odot} / \rm yr$. The mass-loss rates of 
white dwarfs on the cooling sequence (if winds exist at all) are unknown. 

The sdB stars can be identified with models of the extreme horizontal branch 
(EHB) stars with masses of $M_{*} \approx 0.5 M_{\odot}$ (Heber \cite{heb86}; 
Saffer et al. \cite{saf94}), effective temperatures of 
$20000 \, \rm K \leq T_{\rm eff} \leq 40000 \, \rm K$, and surface gravities of 
$5.0 \la \log g \la 6.0$. Helium is usually  deficient with number ratios of   
$10^{-4} \la \rm He / \rm H \la 0.1$ (e.g. Edelmann et al. \cite{edel03}; 
Lisker et al. \cite{lisk05}). However, these helium abundances are 
by at least one order of magnitude more than would be expected from the 
equilibrium condition between gravitational settling and radiative 
levitation (Michaud et al. \cite{mic89}). A possible explanation for this 
discrepancy is the presence of weak winds that retard gravitational 
settling (Fontaine \& Chayer \cite{fon97}; Unglaub \& Bues \cite{ub98}).  
For mass-loss rates $\dot M \approx 10^{-13} M_{\odot} / \rm yr$, 
within the lifetimes of sdB stars near the EHB ($\approx 10^{8} \rm yr$), 
the helium abundances would gradually decrease from the 
solar value to $\rm He / \rm H \approx 10^{-4}$.

The helium deficiencies in sdB stars are accompanied by anomalies in the 
abundances of heavy elements. Some of the most recent results of spectral
analyses are from Geier et al. (\cite{gei08}), O'Toole \& Heber (\cite{otol06}), 
Edelmann et al. (\cite{edel06}), 
Blanchette et al. (\cite{bla06}), and Chayer et al. (\cite{chay06}). The hotter 
sdBs with $T_{\rm eff} > 30000 \, \rm K$ show strong 
deficiencies in many cases especially of light metals like Al, Mg, O, and Si by more than a 
factor of $100$ in comparison to the solar abundances, whereas enrichments 
of elements heavier than iron  
by a factor near $100$ are not unusual. The 
abundances of iron and nitrogen are usually close to the solar value. 
If the scenario with diffusion and chemically homogeneous winds were the 
correct explanation of these compositions, it should be possible to find a 
mass-loss rate for which the anomalies of all elements can be explained 
simultaneously. For the case with solar initial abundance, the calculations 
of Unglaub \& Bues (\cite{ub01}) show that, in the case of chemically 
homogeneous winds, helium should always be more deficient than at least the 
elements C, N, and O, which have been taken into account. No mass-loss rate 
exists that leads to deficiencies in C and O by more than a factor of 
$100$, whereas helium is only deficient by a factor of ten. This, however, 
is not an unusual composition in sdB stars (e.g. Heber et al. 
\cite{heb00}).

Pulsating and non-pulsating sdB stars with similar stellar parameters 
coexist in the $T_{\rm eff}$-$\log g$ diagram (Charpinet et al. 
\cite{char06}). The proposed pulsation mechanism is associated with a local 
enhancement of iron (or other iron group elements) in a mass depth of about 
$10^{-7} M_{*}$, which is indeed expected from the equilibrium condition 
between gravitational settling and radiative levitation (Charpinet et al. 
\cite{char97}; Fontaine et al. \cite{fon03}). However, according to the 
calculations of Chayer et al. (\cite{chay04}) and Fontaine et al. 
(\cite{fon06}), a weak wind with only $\dot M = 6*10^{-15} M_{\odot} / \rm yr$  
would be sufficient to destroy the iron reservoir within a timescale of 
$10^{7} \rm yr$. For higher mass-loss rates, this should happen in even 
shorter time scales. Thus for $\dot M \approx 10^{-13} M_{\odot} / \rm yr$, 
which would be required to explain the helium abundances, this pulsation 
mechanism would become questionable.  

Mass-loss has been detected in sdO stars that are more luminous than sdBs 
(Hamann et al. \cite{ham81}; Rauch \cite{rau93}). 
For sdB stars, up to now there has been no observational proof 
for the existence of winds. From a quantitative analysis of the $\rm H \alpha$ line 
profiles of 40 sdB stars (Maxted et al. \cite{max01}), a comparison of 
synthetic NLTE $\rm H \alpha$
line profiles from static model atmospheres with the observations have revealed 
perfect matches for almost all stars. Only in the four most luminous sdBs have
anomalous $\rm H \alpha$ lines with a small emission at the line centre  
been detected, which are possibly  signatures of weak winds 
(Heber et al. \cite{heb03}).
For the case with $T_{\rm eff} = 36000 \, \rm K$, $\log g = 5.5$, and $\log L/L_{\odot} = 
1.51$,  Vink (\cite{vink04}) found a similar behaviour of $\rm H \alpha$ from 
a spectral analysis of $\rm H \alpha$ with his wind code, if the existence of 
a weak wind with $\dot M \approx 10^{-11} \rm M_{\odot} / \rm yr$ is assumed.
 
For several effective temperatures in the range 
$25000 \, \rm K \leq T_{\rm eff} \leq 50000 \, \rm K$ and for various 
metallicities, mass-loss rates will be 
predicted according to the theory of radiation-driven winds from 
Castor et al. (\cite{cas75}, henceforth CAK). As the usual scaling laws may 
be unreliable for thin winds, the method of solution has been changed slightly 
(see Sect. 2), so it is taken into account that the radiative force is limited 
for small wind optical depth parameters and tends to some maximum value.
To simplify the numerical method, later improvements in the CAK 
theory are neglected. As discussed in Sect. 6, it should be possible to 
derive at least an upper limit for the mass-loss rates.
For each wind model, we 
checked whether a one-component description may be justified (see Sect. 3).
The results for $T_{\rm eff} = 40000$ and $50000 \, \rm K$ and several metallicities, 
which have been obtained with line force multipliers according to 
Kudritzki (\cite{kud02}), are presented in Sect. 4. The results 
for $T_{\rm eff} = 25000$, $30000$, and $35000 \, \rm K$ (Sect. 5) were 
obtained with force multipliers according to our own calculations (see Sect. 2.4.2). 
The predicted mass-loss rates are compared 
with the results of Vink \& Cassisi (\cite{vink02}). 

In Sects. 6.3 and 6.4, we discuss how the abundance anomalies 
in sdB stars could be explained. The arguments are also 
relevant for hot white dwarfs and chemically peculiar main sequence stars, 
which are reviewed by Smith (\cite{smith96}). A subgroup are the HgMn 
stars. Because they are characterised by very low rotational velocities and weak 
or non-detectable magnetic fields, they are one of the best natural 
laboratories for studying the competing processes of gravitational settling and 
radiative diffusion (Vauclair \& Vauclair \cite{vvg82}). The HgMn stars are 
characterised by enhancements of heavy metals (e.g. Hg, Mn, Pt, Sr, Ga), 
deficiencies of some light elements (e.g. He, Al, N), and  
isotopic anomalies of metals. Some of the most recent 
papers about these stars are from Zavala et al. (\cite{zav07}) and 
Adelman et al. (\cite{adel06}). Similar to sdB stars, weak winds with   
mass-loss rates of $10^{-14}$ to $10^{-12} M_{\odot} / \rm yr$ 
could lead to abundance anomalies that are different from the ones 
obtained from the equilibrium condition between 
gravitational settling and radiative levitation
(Landstreet et al. \cite{land98}). However, for the 
typical stellar parameters $10500 \, \rm K \leq T_{\rm eff} \leq 16000 \, \rm K$ 
and $\log g \approx 4.0$, these winds can hardly be chemically homogeneous, 
because the radiative force is too low (Babel \cite{bab96}).

In chemically peculiar main sequence stars, the abundance anomalies 
depend on stellar rotation  (see e.g. Vauclair \cite{vauc03}), on the presence of 
magnetic fields (Turcotte \cite{tur03}) and of convection zones. 
In hydrogen-rich hot white dwarfs and sdB stars, these effects do not seem  
to be the most important 
ones. In general both are slow rotators (Karl et al. \cite{karl05}; 
Koester et al. \cite{koes98}; Heber et al. \cite{heb97}; Heber \& Edelmann 
\cite{heb04}). O'Toole (\cite{otol05}) finds no correlation between 
magnetic field strengths in sdB stars and abundance anomalies. In some 
(pre-) white dwarfs, magnetic fields have been detected 
(Jordan et al. \cite{jor05}; \cite{jor07} and references therein). 
However, diffusion is effective 
in all white dwarfs and not restricted to a subgroup of them.
A thin superficial convection zone with a mass depth of about 
$10^{-12} M_{*}$ may be present in sdB stars, but only for helium abundances 
$\rm He / \rm H \ga 0.01$ by number (Groth et al. \cite{gro85}).    
%=================================================================================
\section{Calculation of mass-loss rates}
%=================================================================================
CAK introduced the dimensionless optical depth parameter
\begin{equation}
t = \sigma_{\rm e} v_{\rm th} \rho \left ( \frac {dv}{dr} \right )^{-1} \, ,
\end{equation}
where $\rho$ is the density and $dv/dr$ is the velocity gradient. 
Here, $\sigma_{\rm e}$ is the electron scattering opacity with 
\begin{equation}
\sigma_{e} = \frac {n_{\rm e}}{\rho} \sigma_{\rm T}
\end{equation}
where $\sigma_{\rm T} = 6.6524 * 10^{-25} \rm cm^{2}$ is  the Thomson cross section for 
electrons and $n_{\rm e}$ the electron number density. The ratio $n_{\rm e} / \rho$ 
depends on the composition and on the degree of ionization. Both are assumed to 
be constant throughout the wind. In addition, the wind is assumed to be isothermal 
with $T = T_{\rm eff}$. Thus the mean thermal velocity $v_{\rm th}$ of hydrogen 
is
\begin{equation}
v_{\rm th} = \sqrt{\frac {2 k_{\rm B} T_{\rm eff}}{m_{\rm H}}}
\end{equation}
where $k_{\rm B}$ is the Boltzmann constant and $m_{\rm H}$ the atomic mass 
of hydrogen. 
The line radiative acceleration can be written in terms of the line force 
multiplier $M \left ( t \right )$ and the luminosity $L$ of the star
\begin{equation}
g_{\rm rad} = \frac {1}{c} \sigma_{\rm e} \frac {L}{4 \pi r^{2}} 
              M \left ( t \right ) \, .
\end{equation}	      
In the original CAK theory, the force multiplier is parametrized 
according to $M \left ( t \right ) = k t^{- \alpha}$. With constant 
force multiplier parameters $k$ and $\alpha$ and if the radiative acceleration 
due to electron scattering is neglected, it follows that
$\dot M \sim L^{\frac {1}{\alpha}}$(see e.g. Puls et al.  \cite{puls00}). Usually 
$\alpha$ is between about $0.5$ and $0.7$. However, the results of such scaling laws 
are questionable for the case of thin winds. 
This is because, according to the parametrized form, the force 
multiplier and thus the radiative force may become arbitrarily strong if the wind optical depth 
parameter is small enough. As a consequence, even for such compact stars as 
white dwarfs, for which in fact no wind solution exists, the scaling laws derived for 
more luminous stars predict small, but non-zero mass-loss rates. 
 
For $t \rightarrow 0$ 
the $\log M \left ( t \right )$ - $\log t$ is not a straight line as would follow 
from the parametrized form, but the relation flattens and the slope 
$- \alpha$ of this relation approaches zero (see Sects. 2.4 and 5.1). 
For given excitation and ionization 
equilibrium, the force multiplier smoothly tends to 
a maximum value $M_{\rm max}$. To take this into account, Kudritzki (\cite{kud02})
developed a method with variable force multiplier parameters to calculate wind 
models for extremely metal-deficient hot stars. 
The methods used in the present paper as described in 
Sects. 2.1 and 2.2 allow the calculation of the mass-loss rate without parametrization 
of the force multiplier. However, in contrast to Kudritzki's method, later 
improvements in the CAK theory like the finite disk correction factor (Pauldrach et al. 
\cite{paul86}; Friend \& Abbott \cite{fri86}) and changes in ionization in the 
wind are neglected.

For fixed excitation and ionization equilibrium, the force multiplier depends on the 
wind optical depth parameter alone (see Sect. 2.4.2). 
With the equation of continuity
\begin{equation}
\dot M = 4 \pi r^{2} \rho v
\end{equation}
and the definition
\begin{equation}
D = r^{2} v \frac {dv}{dr} \, ,
\end{equation}
the wind optical depth parameter can be written as
\begin{equation}
t = \sigma_{\rm e} v_{\rm th} \frac {\dot M}{4 \pi D} \, ,
\end{equation}
so $t$ and thus the force multiplier are a function of $\dot M$ and $D$.
Then, as described below, the solution to the momentum equation is 
straightforward. The inclusion of 
the improvements in the CAK theory would lead to additional dependencies, 
because the finite disk correction factor, as well as the ionization equilibrium, 
change with radius. This may complicate the method of solution.
Moreover, in weak winds these effect are probably less important than the 
effect of the shadowing of the flux by the photospheric lines discussed by 
Babel (\cite{bab96}), which is 
not taken into account in the original CAK theory and thus in the 
present calculations. 
The consequences of the various simplifications and their relative 
importance will be discussed in Sect. 6.    

The momentum equation for a stationary wind is
\begin{equation}
v \frac {dv}{dr} = - \frac {1}{\rho} \frac {dp}{dr} - \frac {G M_{*}}{r^{2}}
\left ( 1 - \Gamma_{\rm e} \right ) + g_{\rm rad}
\end{equation}
where $dp / dr$ is the gradient of the gas pressure, $G$ the gravitational constant, 
and $\Gamma_{\rm e}$ is the Eddington factor:
\begin{equation}
\Gamma_{\rm e} = \frac {L \sigma_{\rm e}}{4 \pi c G M_{*}} \, .
\end{equation}
For sdB stars, $\Gamma_{\rm e}$ is only about $0.01$.
The solution to the momentum equation is especially simple, if the contribution 
of the gas pressure is completely (see Sect. 2.1) or partially (Sect. 2.2) neglected.
%*****************************************************************************
\subsection{Momentum equation without gas pressure}
%*****************************************************************************
If the contribution of the gas pressure to the momentum equation is  
neglected and if the force multiplier is given in parametrized form with 
constant parameters, then analytical 
formulae for the mass-loss rate and terminal velocity can be derived (see Sect. 8.7.1
of Lamers \& Cassinelli, \cite{lam99}). The method is similar for
non-parametrized force multipliers; but 
then the equations must be solved numerically.
If in Eq. (8) $dp / dr$ is 
neglected, Eq. (4) for $g_{\rm rad}$ inserted, and the resulting equation 
multiplied with $r^{2}$, then the momentum equation with the definition of $D$  
can be written as
\begin{equation}
H \equiv D + G M_{*} \left ( 1 - \Gamma_{\rm e} \right )
- \frac {L \sigma_{\rm e}}{4 \pi c} M \left ( t \right ) = 0 \, .
\end{equation}
With the present assumptions, this equation only depends on $\dot M$ and $D$. 
All other quantities are fixed. 
For a given mass-loss rate, it has either two solutions, one or no 
solution for $D$. The mass-loss rate according to the CAK theory corresponds 
to the case with one solution. Then the critical point of CAK degenerates 
and every point in the flow becomes critical. This solution fulfils the 
critical point condition $dH / d (dv/dr) = 0$, which is equivalent to
\begin{equation}
\frac {dH}{dD} = 0 \, .
\end{equation} 
With Eq. (10), it follows that
\begin{equation} 
 1 - \frac {L \sigma_{\rm e}}{4 \pi c} 
 \frac {dM \left ( t \right )}{dt} \frac {dt}{dD} = 0 \, . 
\end{equation} 
Equations (10) and (12) are solved numerically for $\dot M$ and for the critical value 
$D_{\rm c}$. From insertion of $\dot M$ and 
$D_{\rm c}$ into Eq. (7) the critical optical depth parameter $t_{\rm c}$ follows,
which for completely neglected gas pressure is constant throughout the flow. 
The solution $v \left ( r \right )$ is 
obtained by quadrature of Eq. (6) with $D = D_{\rm c}$, so that
\begin{equation}
v \left ( r \right ) = \left ( v_{0}^{2} + \frac {2 D_{\rm c}}{R_{*}}
\left ( 1 - \frac {R_{*}}{r} \right ) \right )^{\frac {1}{2}}
\end{equation}
where $v_{0}$ is the velocity at $r=R_{*}$. For $r \rightarrow \infty$ and with 
$v_{0} \ll v_{\infty}$ the terminal velocity is
\begin{equation}
v_{\infty} = \sqrt {\frac {2 D_{\rm c}}{R_{*}}} \, .
\end{equation}
This gives the usual $\beta$ - type velocity law with $\beta = \frac {1}{2}$. 
%****************************************************************************** 
\subsection{The shooting method}
To neglect the gas pressure completely is only justified in the supersonic region, 
where the wind essentially is driven by the radiative force. If one is interested 
in the solution $v \left ( r \right )$ of the momentum equation in the region around 
the sonic point and in the subsonic region, the gas pressure must at least be partially taken 
into account.
With the equation of continuity
and the equation of state for a perfect gas, the gradient of the gas pressure can 
be written as
\begin{equation}
\frac {dp}{dr} = - \rho \left ( \frac {a^{2}}{v} \frac {dv}{dr} + \frac {2 a^{2}}{r}
                 \right ) \, .
\end{equation}
Here $a$ is the isothermal sound speed. With  $T = T_{\rm eff}$, it is 
\begin{equation}
a = \sqrt {\frac {k_{\rm B} T_{\rm eff}}{\mu}}
\end{equation}
where $\mu$ is the mean particle mass.
The first term in brackets on the righthand side of Eq. (15) corresponds to an 
acceleration caused by the gradient of the gas pressure in planar geometry.
This is the contribution to the momentum equation, which is important in the subsonic 
region near the photosphere of the star. The term $2 a^{2} / r$ is the acceleration due 
to the area expansion. 

In winds with temperatures $T \approx T_{\rm eff}$, the gravitational acceleration 
$g {\left ( r \right )}$ exceeds this curvature term by more than a factor of $100$ 
near the photosphere. As, however, $g { \left ( r \right )}$ 
decreases proportionaly to $r^{-2}$, whereas the curvature term decreases only with 
$r^{-1}$, a radius exists at which both accelerations are equal. This is denoted 
as the Parker point (Lamers \& Cassinelli, \cite{lam99}) and is located far outside the 
star at distances of some hundreds of stellar radii. In a wide region between the 
sonic point and the Parker point, the contribution of the gas pressure to the momentum 
equation is small. As we intend to solve 
the momentum equation by outward integration from the photosphere, at least the  
first part of $dp / dr$, which is important in the subsonic region, must be 
taken into account. The curvature term, however, will be neglected. Then 
with Eqs. (8) and (15) it follows that    
\begin{equation}
v \frac {dv}{dr} \left ( 1 - \frac {a^{2}}{v^{2}} \right ) 
= - \frac {G M_{*}}{r^{2}} \left ( 1- \Gamma_{\rm e} \right ) + g_{\rm rad} \, .
\end{equation}
In the complete momentum equation the term $2 a^{2} / r$ should appear on the righthand 
side. The omission of this term simplifies the solution topology and the numerical 
method (see Sect. 2.3).  
As the density in the photosphere is known from the 
hydrostatic equation, the velocity $v_{0}$ at the 
inner boundary (at $r=R_{*}$, which is assumed at a Rosseland mean optical depth 
$\overline \tau = 2/3$) can be obtained for a trial value of $\dot M$ 
from the equation of continuity. With this initial value, 
Eq. (17) is integrated outwards. The highest mass-loss rate, for which a solution 
can be extended from the photosphere to infinity, is the one according to the 
CAK theory.  
For all cases discussed in the present paper, the mass-loss rates have been calculated from 
both methods described in Sects. 2.1 and 2.2. The results are very similar, because the 
differences do not exceed numerical uncertainties.
%*************************************************************************************
\subsection{Example: O5 main sequence star}
%*************************************************************************************
To illustrate the methods described in the previous sections, we now calculate a wind model 
for an O5 main sequence star using the same stellar parameters as CAK:
\begin{displaymath}
T_{\rm eff} = 49290 \, \rm K \, , \qquad \log g = 3.94 \, , \qquad M_{*} = 60 \, M_{\odot} \, ,
\end{displaymath}
and $L = 9.66 *10^{5} L_{\odot}$, $\Gamma_{\rm e} = 0.4$, $\sigma_{\rm e} = 0.325 
\, \rm cm^{2} \, \rm g^{-1}$.
As in CAK for the force multiplier it is assumed that
\begin{equation}
M \left ( t \right ) = \frac {1}{30} \left ( \frac {1}{\sqrt {12}} t \right )^{-0.7} \, .
\end{equation}
The factor $1 / \sqrt {12}$ appears because CAK used the optical depth parameter for 
a carbon ion, so that their $v_{\rm th}$ is the thermal velocity of carbon.
Although $M \left ( t \right )$ is given in parametrized form, we do not make use of 
the parametrization during the calculations.

\begin{figure}
\centering
\includegraphics[width=7.5cm]{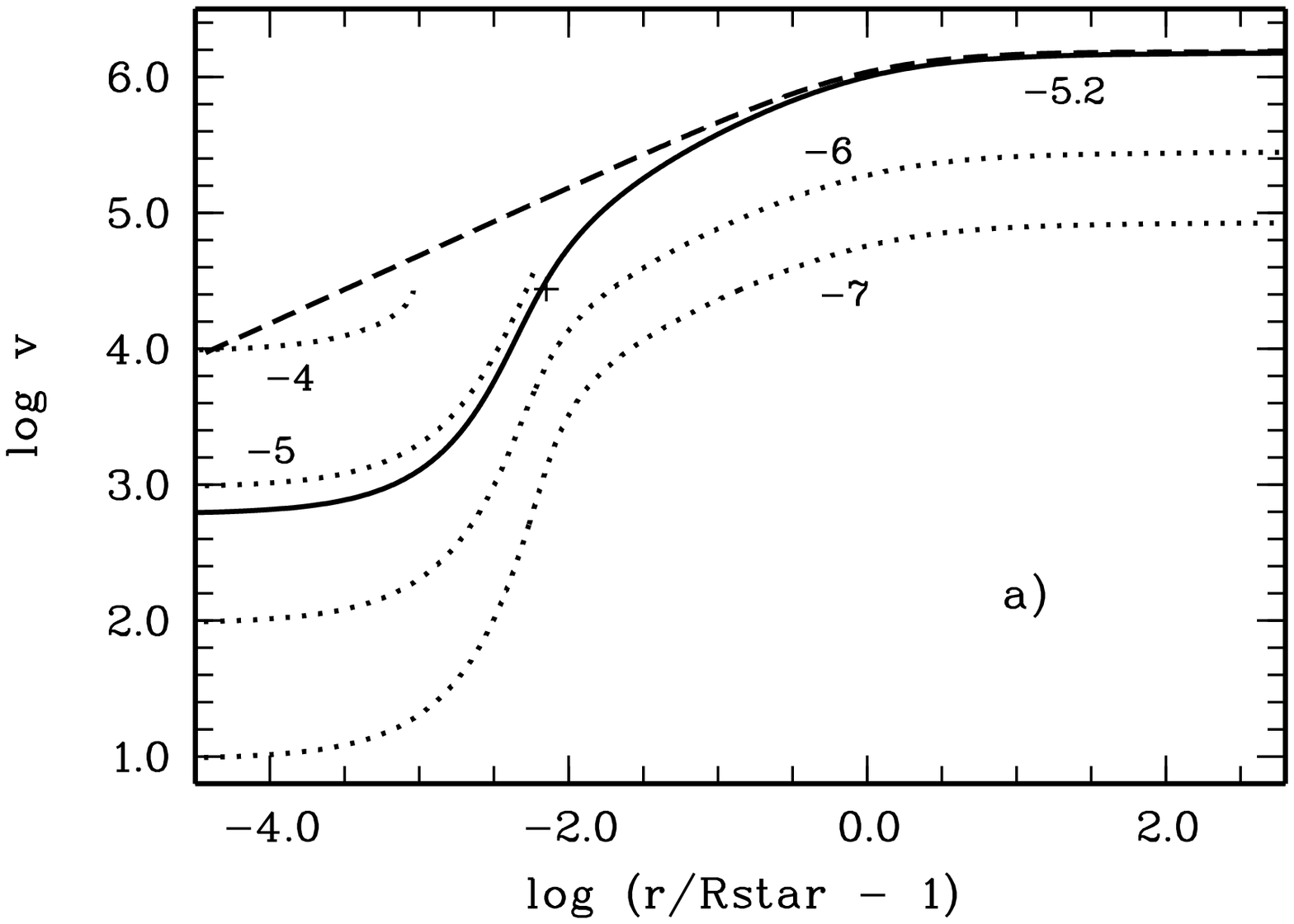}
\\[1.0cm]
\includegraphics[width=7.5cm]{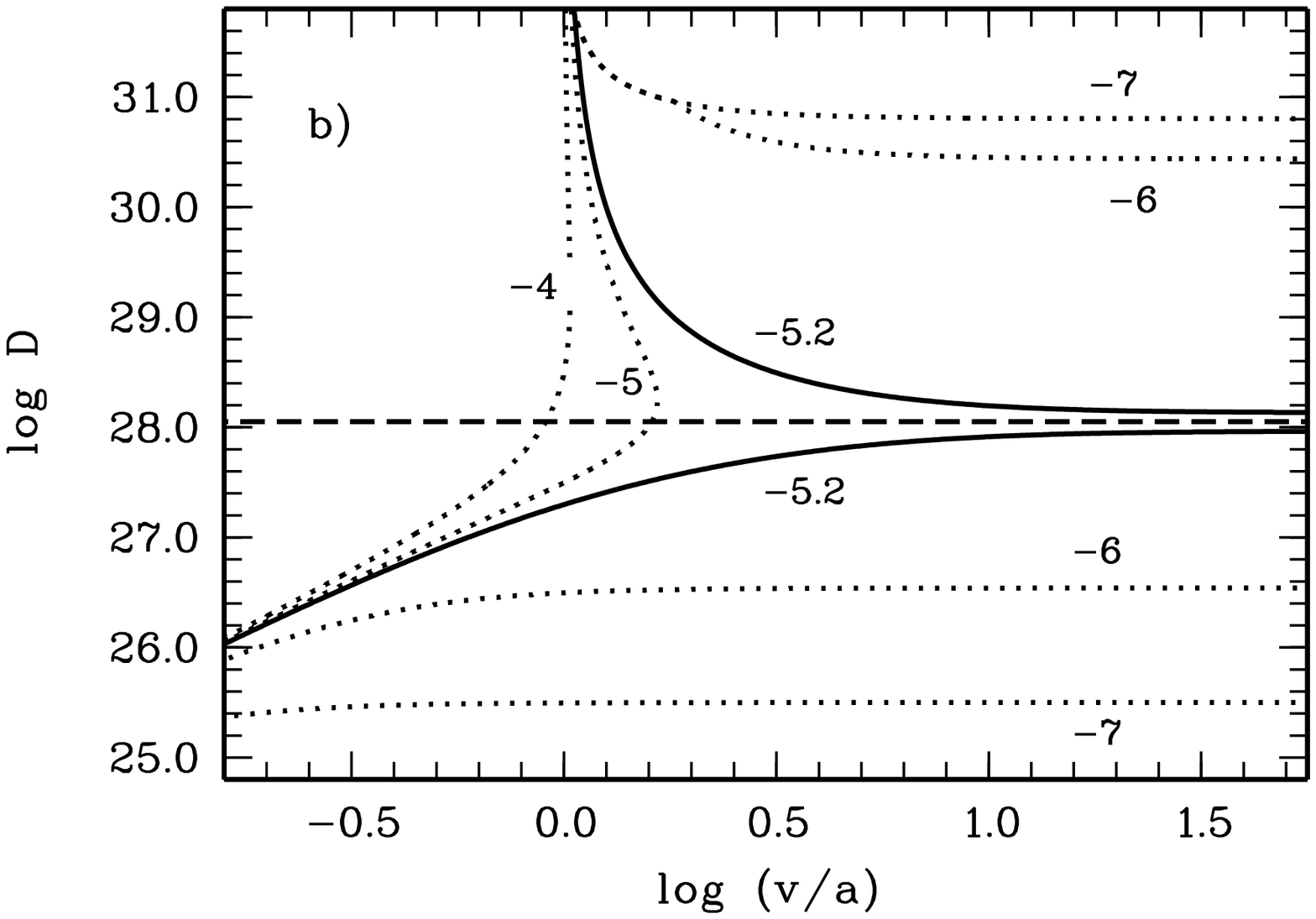}
\\[1.0cm]
\caption{Results for 
$T_{\rm eff} = 49290 \, \rm K$, $\log g = 3.94$, $M_{*} = 60 \, M_{\odot}$. 
a) Solutions $v \left ( r \right )$ (in $\rm m \, \rm s^{-1}$) obtained with the shooting method 
for $\dot M = 6.4 * 10^{-6} M_{\odot} / \rm yr$ (solid line; the cross indicates the 
location of the sonic point), $\dot M = 10^{-4}$, $10^{-5}$, 
$10^{-6}$ and $10^{-7} M_{\odot} / \rm yr$ (dotted lines). The dashed line is the 
critical solution of the momentum equation with completely neglected gas 
pressure. b) Solutions D (in $\rm cm^{3} \, \rm s^{-2}$) of the momentum equation (17) 
as a function of the velocity for the mass-loss rates as in Fig. 1a. The horizontal dashed 
lines represents the critical value $D_{\rm c}$ obtained from the solution of the 
momentum equation with completely neglected gas pressure. In both figures the curves 
are labelled with $\log \dot M$.}
\end{figure}
Solutions for $\dot M = 10^{-4}$ and $10^{-5} \rm M_{\odot} / \rm yr$  
fail at velocities somewhat higher than the sound speed. Thus these mass-loss rates are 
too high. The highest mass-loss rate for which the solution can be extended outwards to 
infinity and the corresponding terminal velocity are
\begin{displaymath}
\dot M = 6.4 * 10^{-6} \, \frac {\rm M_{\odot}}{\rm yr} \qquad 
v_{\infty} = 1527 \, \frac {\rm km}{\rm s} \, .
\end{displaymath}
For comparison, CAK obtained 
$\dot M = 6.6*10^{-6} \rm M_{\odot} / \rm yr$ and $v_{\infty} = 1515 \, \rm km / \rm s$. 
If the finite disk correction factor were taken into account and all other assumptions 
are unchanged, then Pauldrach et al. (\cite{paul86}) obtain
$\dot M = 3.5*10^{-6} M_{\odot} / \rm yr$ and $v_{\infty} = 5123 \, \rm km / \rm s$.
After multiplication of Eq. (17) with $r^{2}$, with Eq. (4) for $g_{\rm rad}$, it follows 
that with the present assumptions Eq. (17) 
depends on $\dot M$, $D$, and $v$. In the limit $v \gg a$, it is identical 
with the momentum equation with completely neglected gas pressure. Thus for high 
velocities the solutions of Eqs. (10) and (17) approach each other, whereas in 
the inner regions, where the gas pressure 
is essential, the differences are great, as can be seen from comparing the solid and 
dashed lines in Fig. 1a.
In addition to the critical solution, shallow solutions also exist. They are 
characterised by lower mass 
loss rates, e.g. $\dot M = 10^{-6}$ or $10^{-7} M_{\odot} / \rm yr$ 
and terminal velocities below the CAK value.

In Fig. 1b the solutions $D$ of Eq. (17) are shown as a function of the velocity. In 
the subsonic region one solution for $D$ (and thus the velocity gradient) exists.
In the supersonic region, however, two solutions exist for each mass-loss rate: 
a shallow solution with a low value of D and a steep one with a high value of $D$.
For mass-loss rates higher than the critical value, the shallow and steep solution 
meet, so that for velocities somewhat higher than $a$, no solution of the momentum 
equation exists any more. The critical value of $\dot M$ is the highest mass-loss rate for 
which both the shallow and the steep solution can be extended outwards. 
These solutions approach each other and for $v \gg a$ they converge to the critical 
solution of the momentum equation with completely neglected gas pressure (in this case 
$D$ is constant). For subcritical mass-loss rates, the shallow and the corresponding steep 
solution converge to different values of $D$ for $v \gg a$.

The solution topology of the complete momentum equation with the curvature term taken 
into account is more complicated (Bjorkman \cite{bjo95}). It has an X-type topology 
in $D$ and several critical points exist. Then the shallow 
solutions meet at the Parker point and cannot be continously (in D) extended outwards. 
For this reason they are ruled out by CAK. Then for only one 
mass-loss rate a smooth and 
continuous solution exists, which reaches from the photosphere to infinity. To find this 
solution, it is necessary to switch from a shallow to a steep solution at the 
CAK critical point. As in the present case with neglected curvature term shallow solutions can 
also be extented from the photosphere to infinity, only 
an upper limit can be set for the mass-loss rate from the time-independent equations. 
From time-dependent calculations, however, 
Feldmeier \& Shlosman (\cite{feld02}) suggest that shallow solutions evolve in time towards 
higher velocities and mass-loss rates. Thus their physical relevance is questionable.  

As the steep solutions do not exist in the subsonic region, the outward integration of the 
momentum equation from the photosphere must start on a shallow solution. As for the critical 
mass-loss rate, the shallow and steep solution are identical in the limit $v \gg a$ (which 
would not be the case if the curvature term were taken into account), the critical solution 
can be found without switching from the shallow to 
the steep solution. This simplifies the numerical 
method, especially in the case of weak winds, in which the wind optical parameter is in a range 
where the $\log M \left ( t \right )$ - $\log t$ relation is curved. 
Due to the X-type solution topology of the complete momentum equation, the integration  
should start at the critical point in this case. This is the method used by CAK. To set the 
initial conditions at the CAK critical point, the knowledge of the slope of the
$\log M \left ( t \right )$ - $\log t$ relation (which corresponds to the force multiplier 
parameter $\alpha$) is required. Therefore this method leads to complications, if this relation 
is curved and thus $\alpha$ is not a constant.

As can be seen from Fig. 1b, a solution with a subcritical mass-loss rate and a terminal 
velocity higher than the CAK value would require a jump in $D$ from a shallow onto a steep 
solution somewhere in the supersonic region (higher values of $D$ lead to higher terminal 
velocities as can be seen from quadrature of Eq. (6)). To see if a discontinuity in $D$ 
may be physically 
reasonable, the time-dependent equations should be considered. As, however, the CAK theory is 
based on the use of the Sobolev approximation for calculating the radiative force, 
the mathematical nature of the momentum equation would change if the Sobolev approximation 
were dropped (Lucy \cite{lu07a}; \cite{lu07b}). As the Sobolev approximation is questionable 
at least in the subsonic region, it cannot be excluded that smooth and 
continuous solutions with subcritical mass-loss rates and high terminal velocities indeed exist
in this case.
%**************************************************************************
\subsection{The force multiplier}
%**************************************************************************
The results for $T_{\rm eff} = 40000 \, \rm K$ and $50000 \, \rm K$, which will be 
presented in Sect. 4, were
obtained with force multipliers according to Kudritzki (\cite{kud02}). To
avoid the methods of solution described in Sects. 2.1 and 2.2 being 
complicated by changing ionization and excitation equilibrium,  
the dependence of Kudritzki's force multipliers on the degree of ionization is only 
approximately taken into account. How this is done, is described in 
Sect. 2.4.1. 

The results for $T_{\rm eff} = 25000$, $30000$, and $35000 \, \rm K$, 
which will be presented in Sect. 5, have been obtained with force multipliers 
from own calculations. These calculations and the various assumptions are 
described in Sect. 2.4.2.
%+++++++++++++++++++++++++++++++++++++++++++++++++++++++++++++++++++++
\subsubsection{Force multipliers according to Kudritzki (\cite{kud02})}
%++++++++++++++++++++++++++++++++++++++++++++++++++++++++++++++++++++++
For $T_{\rm eff} = 40000$, $50000$, and $60000 \, \rm K$, Kudritzki (\cite{kud02}) 
calculated force multipliers 
with a line list of $2.5*10^{6}$ lines of $150$ ionic species and with approximate 
non-LTE occupation numbers as a function of the wind optical depth parameter and 
of $n_{\rm e} / W$, where $n_{\rm e}$ is the electron density and $W$ the 
geometrical dilution factor ($W \approx 0.5$ near the photosphere).
\begin{figure}
\centering
\includegraphics[width=7.5cm]{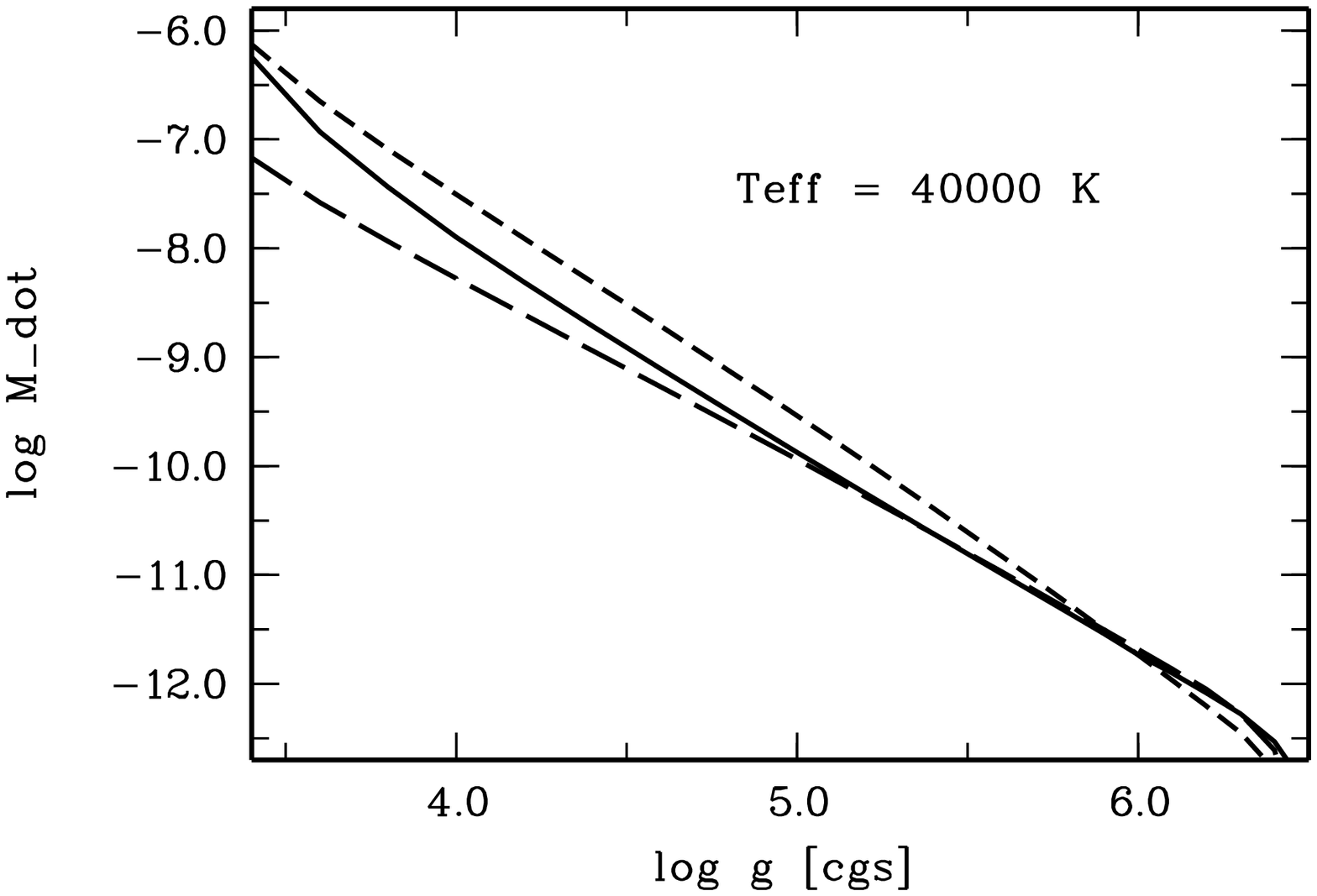}
\\[1.0cm]
\includegraphics[width=7.5cm]{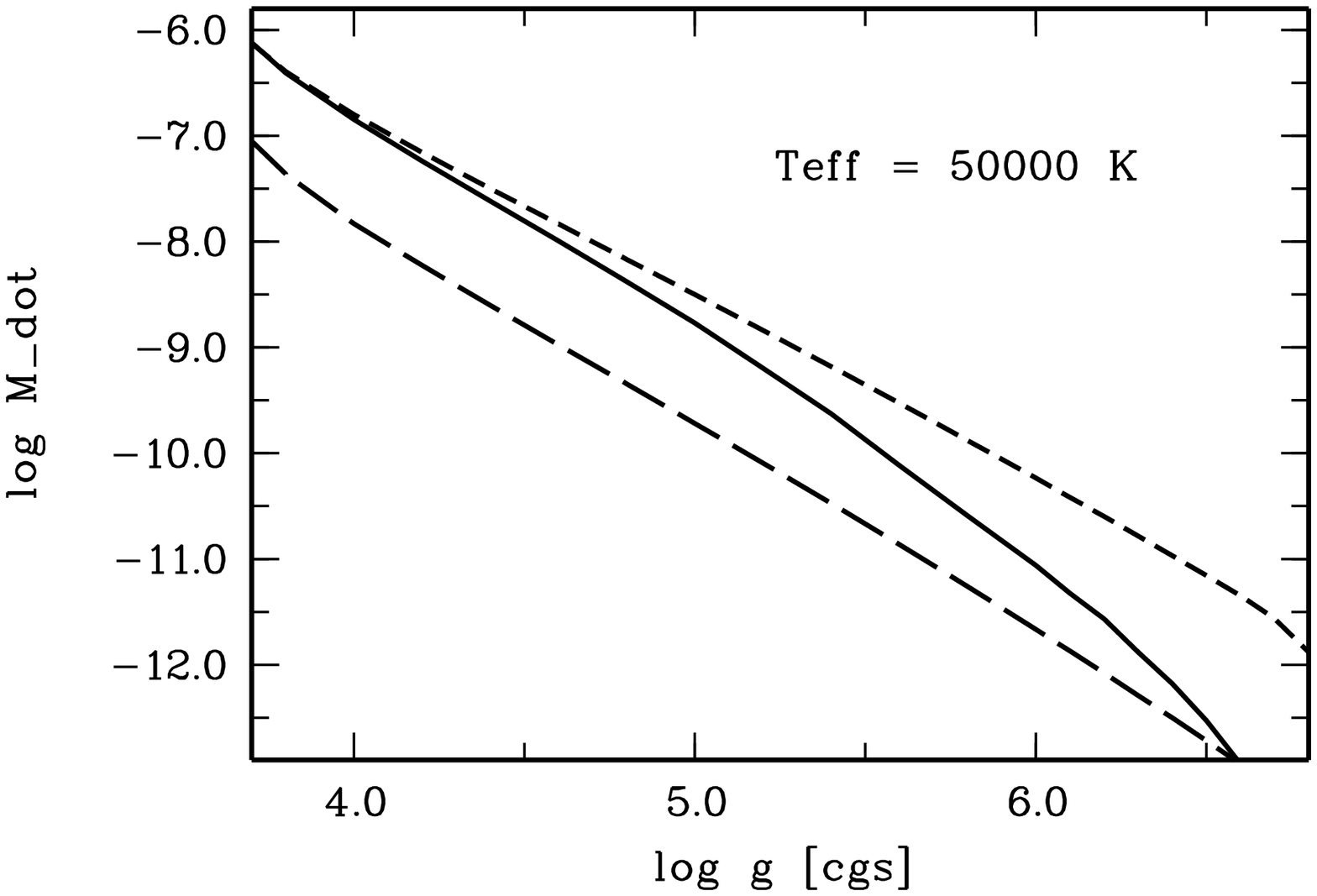}
\\[1.0cm]
\caption{Mass-loss rates (in $M_{\odot} / \rm yr$) as a function of surface 
gravity for $T_{\rm eff} = 40000 \, \rm K$, $50000 \, \rm K$, and $Z/Z_{\odot} = 1$ 
calculated with 
force multipliers according to Kudritzki (\cite{kud02}). The solid lines represent 
the results obtained with the iteration procedure for $n_{\rm e} / W$ as described in 
the text. The short-dashed and long-dashed lines represent the results for  
$n_{\rm e} / W = 10^{14}$ and $10^{10} \rm cm^{-3}$, respectively.}
\end{figure}
In the present calculations $n_{\rm e} / W$ is assumed to be fixed throughout the wind. 
The dependence on the degree of 
ionization is taken into account with the following iteration procedure. 
A first wind model is calculated with an input value  
(e.g. $n_{\rm e} / W = 10^{11} \rm cm^{-3}$). 
From this model we obtain the mean value of $n_{\rm e} / W$ at the sonic 
point and at a radius $r = 2 R_{*}$. The iteration is complete, if this output value is 
consistent with the input value within an accuracy of about $10$ \%. 
Otherwise the iteration is repeated with an improved input value
to ensure that the value of $n_{\rm e} / W$ used in the calculation 
of the force multipliers has a similar order of magnitude to the value of 
the corresponding wind model near the sonic point 
and the CAK critical point (which according to the original theory with the 
complete momentum equation is near $r = 1.5 R_{*}$). 

In Fig. 2 for $T_{\rm eff} = 40000$ and $50000 \, \rm K$ and for
$Z / Z_{\odot} = 1$, the mass-loss rates calculated with this iteration procedure 
are compared to the results obtained with the assumption that it is always 
$n_{\rm e} /W = 10^{14} \rm cm^{-3}$ or $10^{10} \rm cm^{-3}$, respectively, 
regardless how high the wind density is.
It can be seen that, for 
$T_{\rm eff} = 40000 \, \rm K$ and $\log g \ga 5.0$, the results are 
approximately the same in each case with deviations less than a 
factor of two. This is because the dependence of 
the force multiplier on $n_{\rm e} / W$ is weak for small wind optical depth 
parameters, which are expected in thin winds (see Fig. 3). 
The situation is different 
for $T_{\rm eff} = 50000 \, \rm K$. Here the mass-loss rates derived for 
the two values of $n_{\rm e} / W$ may differ by almost two orders of 
magnitude. The values obtained with the iteration procedure are 
in between these results (the finally adopted values of $n_{\rm e} / \rm W$ 
can be read off from Fig. 5 in Sect. 4). 

For $T_{\rm eff} = 60000 \, \rm K$, the dependence of the mass-loss rates 
on the degree of ionization is even stronger;  
e.g., for $\log g = 6.0$ it follows that 
$\log \dot M = -11.4$ with $n_{\rm e} / W = 10^{10} \rm cm^{-3}$ and 
$\log \dot M = -7.6$ with $n_{\rm e} / W = 10^{14} \rm cm^{-3}$. Thus the

predicted value of $\dot M$ depends almost linearily on 
$n_{\rm e} / W$. Then the iteration procedure for $n_{\rm e} / W$ 
does not converge. For this reason the case $T_{\rm eff} = 60000 \, \rm K$ 
is not considered in the present paper.

The force multipliers according to Kudritzki's Eq. (25) have been not allowed to 
exceed a value of $M_{\rm max} = 2000 \, Z / Z_{\odot} + M_{\rm H, \rm He}$, 
according to his Eq. (12). The contribution $M_{\rm H, \rm He}$ of hydrogen 
and helium to $M_{\rm max}$ was estimated from own calculations as 
described below. The adopted values are $M_{\rm H, \rm He} = 263$ for 
$T_{\rm eff} = 40000 \rm K$ and $M_{\rm H, \rm He} = 93$ for 
$T_{\rm eff} = 50000 \rm K$. At least for $T_{\rm eff} = 40000 \rm K$, 
$Z / Z_{\odot} = 1$ this restriction is in good agreement with the 
value $M_{\rm max} = 2340$ obtained from own calculations (see Fig. 3). 
It only affects the results for 
extremely weak winds with wind optical depth parameters 
$\log t \la -7.0$. In these cases decoupling of the ions from hydrogen 
and helium is expected, so the corresponding 
wind models are of questionable physical relevance. Thus this restriction 
does not change the conclusion of the present paper.
%++++++++++++++++++++++++++++++++++++++++++++++++++++++++++++++++++++++
\subsubsection{Force multipliers from own calculations}
%++++++++++++++++++++++++++++++++++++++++++++++++++++++++++++++++++++++   
The results presented in Sect. 5 are obtained with force 
multipliers from own calculations. The elements H, He, C, N, and O 
are taken into account with a line list that is essentially similar to the one 
used in the diffusion calculations of Vauclair et al. (\cite{vvg79})
and later in our diffusion calculations (Unglaub \& Bues, \cite{ub96}).
It consists of about $150$ preferably strong lines with atomic data from 
Wiese et al. (\cite{wie66}). Similar to the diffusion calculations, multiplets, 
or doublets are lumped together into one line (note, however, that this may 
underestimate the radiative force, if not all lines are optically thin). 
All ionization stages are taken into account, with the exception 
of the neutral and singly ionized stage of the CNO elements.
The occupation numbers are calculated with the assumption of LTE, 
for a temperature 
$T = T_{\rm eff}$ and an electron density $n_{\rm e} = 2.0 * 10^{14} \rm cm^{-3}$, 
which is typical of regions near the wind base. As the typical electron  
densities in the outer parts of weak winds are considerably lower and because
the assumption of LTE is unreliable in the wind region, an agreement of the 
derived mass-loss rates with the results of more sophisticated calculations 
can be expected only if the dependence of the radiative acceleration on the 
ionization and excitation equilibrium is weak enough.
\begin{figure}
\centering
\includegraphics[width=7.5cm]{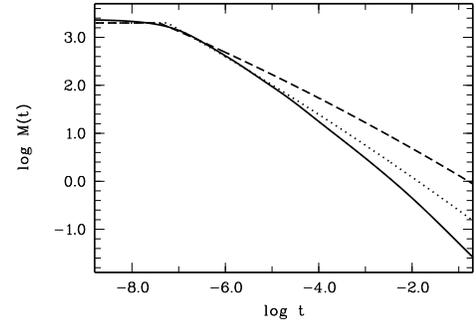}
\\[1.0cm]
\caption{Line force multipliers for $T_{\rm eff} = 40000 \, \rm K$, $Z/Z_{\odot} = 1$ 
as a function of the wind optical depth parameter according to own calculations 
(solid line) and according to Eq. (25) of Kudritzki (\cite{kud02}) for 
$n_{\rm e} / W = 10^{14}$ (dashed line) and $10^{10} \rm cm^{-3}$ (dotted line).
Kudritzki's values have been not allowed to exceed $\log M_{\rm max} = 3.37$.}
\end{figure}

As in Abbott (\cite{abb82}) the force multiplier can be written as
\begin{equation}
M \left ( t \right ) = \sum_{\rm lines} \frac {F_{\nu}}{F} \Delta \nu_{\rm D}
\frac {1 - \exp {\left ( - \eta_{l} t \right )}}
       {t} \, ,
\end{equation}
the sum is over all lines. 
If $\nu_{0}$ is the line centre frequency, the Doppler width is defined here
according to
\begin{equation}
\Delta \nu_{\rm D} = \frac {\nu_{0}}{c}  v_{\rm th} \, .
\end{equation}
Note that $v_{\rm th}$ is the thermal velocity of hydrogen defined in Eq. (3). 
With expressions (1) and (22) for $t$ and $\eta_{l}$, it can be seen that in Eq. (19) 
$v_{\rm th}$ cancels out. 
 
The ratio $F_{\nu} / F$ indicates the monochromatic to total flux at the frequency of the 
line. This flux is calculated from similar 
models as used in the diffusion calculations 
e.g. in Unglaub \& Bues (\cite{ub01}). The temperature structure is obtained with the assumption of 
LTE and with the diffusion approximation for the stellar flux:
\begin{equation}
T^{4} = \frac {3}{4} T_{\rm eff}^{4} \overline \tau + T_{0}^{4}
\end{equation}
with $T_{0}^{4} = \frac {1}{2} T_{\rm eff}^{4}$ at the outer boundary. Here, 
$\overline \tau$ is the Rosseland mean optical depth. The Rosseland mean opacity 
is calculated from the monochromatic continuum opacities for $264$ wavelengths 
in the range $0.1 \AA \leq \lambda \leq 10000 \AA$. Similar to the calculation of 
the force multipliers, the elements H, He, C, N, and O are taken into account, while
$F_{\nu}$ is the emergent monocromatic flux of these models at the frequency of 
the line. It is important to note 
that only the continuum opacity is taken into account and not the opacity due to 
lines. Thus it is assumed that each line in the wind ``sees" the continuum flux, 
which is independent of the velocity and thus the Doppler shift. Although this assumption 
is usual in the CAK theory, it is crucial for the case of thin winds (see Sect. 6.2).

If stimulated emissions are neglected, then $\eta_{l}$ is
\begin{equation}
\eta_{l} \approx \frac {\pi e^{2}}{m_{\rm e} c} f 
                  \frac {n_{i}}{\rho \sigma_{\rm e} \Delta \nu_{\rm D}} 
\end{equation}  
where $e$ is the electron charge, $m_{\rm e}$ the electron mass, $f$ the oscillator 
strength of the line, and $n_{i}$ the number density of particles in the lower level.
With Eq. (2) for $\sigma_{\rm e}$, it can be seen that, for fixed degree of 
ionization and occupation numbers, the quantity $\eta_{l}$ for each line is 
constant throughout an isothermal wind. As in addition the flux $F_{\nu}$ is 
fixed for each line, with these assumptions the force multiplier depends only on $t$.

For $t \rightarrow 0$ the force multiplier converges to its maximum value
\begin{equation}
M_{\rm max} = \sum_{\rm lines} \frac {F_{\nu}}{F} \Delta \nu_{\rm D} \eta_{l} \, .
\end{equation}
With expression (22) for $\eta_{l}$ it can be seen that,
in this optical thin limit,  
the strong lines of the most abundant ions preferably contribute to the 
radiative acceleration.
According to Abbott (\cite{abb82}) for small wind optical depth parameters 
$\log t \la -6$ about $90$ \% of the total force multiplier originates from only 
about $100$ strong lines. Thus for the calculation of extremely weak winds the 
small number of lines taken into account in the own force multiplier calculations 
may be sufficient. For higher wind
densities, however, the radiative acceleration and thus the mass-loss rate 
are underestimated, because in the optically thick limit the contribution of a 
line is independent of its oscillator strength (e.g. Puls et al. 
\cite{puls00}). Then the number of lines taken into account is especially 
important.

In Fig. 3 for $T_{\rm eff} = 40000 \, \rm K$ and solar metallicity, the force 
multipliers from own calculations are compared with the ones according to 
Eq. (25) of Kudritzki (\cite{kud02}) for $n_{\rm e} / W = 10^{14}$ and 
$10^{10} \rm cm^{-3}$, respectively. For the higher values of $t$, the line 
list and the number of elements taken into account in the own calculations is 
not sufficient. Thus the corresponding force multipliers are smaller than 
Kudritzki's ones. For low values of $t$ the results are in good agreement. 
This shows that the force multipliers obtained with the present assumptions 
may agree with the results from more sophisticated calculations if the wind 
optical depth parameter is low enough. However, as the assumption of LTE 
in the present calculations may be unreliable, this agreement also requires  
the dependence on the ionization and excitation equilibrium to be 
small. This is obviously the case in the present example, because 
the differences between Kudritzki's results for $n_{\rm e} / W = 10^{14}$ and 
$10^{10} \rm cm^{-3}$ decrease to lower values of $t$. 
\section{The decoupling of elements}
%**********************************************************************************
According to our calculations, the mean radiative acceleration of the metals 
is by at least two orders of magnitudes higher than of hydrogen and helium.  
This may justify lumping hydrogen and helium together into one ``element" 
(henceforth element 1) and the metals into ``element" 2, although the 
velocities of the individual metals are different (Krti\v cka \cite{krt06}).
 
A chemically homogeneous, radiatively driven wind may only exist 
if the coupling due to Coulomb collisions between metals and hydrogen 
and helium is efficient. If a wind model is calculated from a 
one-component description it is implicitely assumed that this is the case.
Whether this assumption is justified or not may be checked by considering  the 
momentum equations of the individual constituents.   
%*********************************************************************************
\subsection{Equations for an isothermal three-component model}
%*********************************************************************************
For each of the constituents, a momentum equation must be valid:
\begin{equation} 
\rho_{1} v_{1} \frac {dv_{1}}{dr} + \frac {dp_{1}}{dr} + \rho_{1} 
 \frac {G M_{*}}{r^{2}}
- n_{1} Z_{1} e E - \rho_{1} g_{\rm rad}^{\left ( 1 \right )} = \Delta Q_{1}
\end{equation} 
\begin{equation}
\rho_{2} v_{2} \frac {dv_{2}}{dr} + \frac {dp_{2}}{dr} + \rho_{2} 
\frac {G M_{*}}{r^{2}}
- n_{2} Z_{2} e E - \rho_{2} g_{\rm rad}^{\left ( 2 \right )} = \Delta Q_{2} \, .
\end{equation}
Here, $\rho_{1}$, $\rho_{2}$, $n_{1}$, $n_{2}$, and $v_{1}$, $v_{2}$ 
are the mass densities, 
particle densities, and mean velocities of ``elements" 1 (H and He) and 2 
(metals), respectively, and $g_{\rm rad}^{\left ( 1 \right )}$ and 
$g_{\rm rad}^{\left ( 2 \right  )}$
are the mean radiative accelerations and $Z_{1} e$, $Z_{2} e$  the mean 
charges. And $\Delta Q_{1}$ and $\Delta Q_{2}$ are the momentum per unit volume 
and unit time, which is exchanged between the two constituents via collisions. 
Because of momentum conservation, it is $\Delta Q_{1} + \Delta Q_{2} = 0$.
The Coulomb interaction of the elements with the electrons is taken into account 
via the polarization electric field $E$, which is obtained from the momentum 
equation for the electrons:
\begin{equation}
\rho_{\rm e} v_{\rm e} \frac {dv_{\rm e}}{dr} + \frac {dp_{\rm e}}{dr} 
+ \rho_{\rm e} \frac {G M_{*}}{r^{2}} 
+ n_{\rm e} e E - \rho_{\rm e} g_{\rm rad}^{\left ( \rm e \right )} 
 = 0
\end{equation}
where $g_{\rm rad}^{\left ( \rm e \right )}$ is the radiative acceleration 
on the electrons due to Thomson scattering.

The velocities of the various constituents 
are related to the centre of mass velocity $v$ according to
\begin{equation}
v = \frac {\rho_{1} v_{1} + \rho_{2} v_{2} + \rho_{\rm e} v_{\rm e}} 
          {\rho} \, .
\end{equation}
In addition we demand that there is no net electric current so that 
the plasma remains neutral on a macroscopic scale:
\begin{equation}
Z_{1} e n_{1} v_{1} + Z_{2} e n_{2} v_{2} - e n_{\rm e} v_{\rm e} = 0 \, .
\end{equation}
This system of equations, together with the equations of continuity for each 
constituent, and the energy equations (if the wind is not assumed to be 
isothermal) have to be solved  to obtain a multicomponent wind model. This 
has been the approach of Krti\v cka \& Kub\'at (\cite{krt00}, \cite{krt01a}). 
In the present paper for each wind model obtained from the one-component 
description (from which $\dot M$ and 
$v \left ( r \right )$ is known), we check whether the momentum 
transferred from the metals to hydrogen and helium can be sufficient (see Sect. 3.3). 
Possible heating processes (e.g. frictional heating) are neglected,   
we assume always $T = T_{\rm eff}$.
%**************************************************************************
\subsection{The collisional acceleration on H and He}
%**************************************************************************
A mean collisional acceleration on hydrogen and helium due to collisions with 
metals may be defined according to
\begin{equation}
g_{\rm coll}^{\left ( 1 \right )} = \frac {1}{\rho_{1}} \Delta Q_{1} \, .
\end{equation} 
From the equations derived by Burgers (\cite{bur69}) it follows that
\begin{equation}
g_{\rm coll}^{\left ( 1 \right )} = \rho_{2} k_{12} G \left ( x \right )
\end{equation}
with
\begin{equation}
k_{12} = \frac {1}{m_{1} m_{2}} \frac {4 \pi Z_{1}^{2} Z_{2}^{2} e^{4}} {k_{\rm B} T}
\ln \Lambda  \, .
\end{equation}
With the assumed number ratio $\rm He / \rm H = 0.1$, the mean mass of H and He is 
$m_{1} = 1.27 m_{\rm p}$. The 
metals are represented by the CNO elements with abundances proportional to 
solar ones (the solar abundances are from Grevesse \& Sauval \cite{grev98}). 
For the mean mass of the metals, $m_{2} = 14.6 m_{\rm p}$ is assumed. 
The Coulomb logarithm $\ln \Lambda$ is defined according to 
Burgers (\cite{bur69}):
\begin{equation}
\ln \Lambda = - \frac {1}{2} + \ln \left ( \frac {3 k_{\rm B} T R_{\rm D}}{Z_{1} Z_{2} 
               e^{2}} \right ) \, .
\end{equation}
Thus, in an isothermal wind with constant ionization, $k_{12}$ is almost constant apart 
from a weak dependence on density via the Debye radius $R_{\rm D}$, which appears in the 
Coulomb logarithm:
\begin{equation}
R_{\rm D} = \left ( \frac {k_{\rm B} T}{4 \pi e^{2} \left ( n_{1} Z_{1}^{2} + 
n_{2} Z_{2}^{2} + n_{\rm e} \right )} \right )^{0.5}  \, .
\end{equation}
As the particle densities decrease in an outward direction, the Debye radius and thus 
$\ln \Lambda$ increase.
The Chandraskhar function
\begin{equation}
G \left ( x \right ) = \frac {1}{2x^{2}} \left ( \rm erf \left ( x \right ) 
-x \frac {2}{\sqrt {\pi}} \exp \left ( - x^{2} \right ) \right )
\end{equation}  
depends on the 
difference in the mean velocities of the constituents. For $v_{2} > v_{1}$ it is
\begin{equation}
x = \frac {v_{2} - v_{1}}{\alpha}
\end{equation}
with
\begin{equation}
\alpha = \sqrt {\frac {2 k_{\rm B} T}{m_{12}}}  \, ,
\end{equation}
where $\alpha$ has the dimension of a velocity and should not 
be confused with the force multiplier parameter, and
$m_{12} = m_{1} m_{2} / \left ( m_{1} + m_{2} \right )$ is the reduced mass. 
As $m_{12} = 1.17 m_{\rm p}$ in the present case, the quantity $x$ approximately is 
the velocity difference between both constituents in units of the thermal velocity 
$v_{\rm th}$ of hydrogen.
The error function $\rm erf \left ( x \right )$ has been evaluated with the routines 
in Press et al. (\cite{pre92}). For $x \ll 1$ the function 
$G \left ( x \right )$ increases linearly with $x$. For $x = 0.968$, however, it 
reaches its maximum value
\begin{equation}
G_{\rm max} = 0.214
\end{equation}
and decreases to higher values of $x$.
%**************************************************************************
\subsection{Criterion for the existence of a coupled wind}
%**************************************************************************
In a chemically homogeneous wind with a total mass-loss rate $\dot M$, the 
mass-loss rates of the various constituents are by definition (see 
Sect. 1) 
$\dot M_{1} = \zeta_{1} \dot M$ for hydrogen and helium and 
$\dot M_{2} = \zeta_{2} \dot M$ for the metals. Then the equations 
of continuity for both constituents can be written as
\begin{equation}
\zeta_{1} \dot M = 4 \pi r^{2} \rho_{1} v_{1}
\end{equation}
\begin{equation}
\zeta_{2} \dot M = 4 \pi r^{2} \rho_{2} v_{2}  \, .
\end{equation}
The mass fractions $\zeta_{1}$ and $\zeta_{2}$ of hydrogen and helium and 
metals specify the 
metallicity of the star. For solar composition we use $\zeta_{2} = 0.0133$. 
As $\dot M$ is the total mass-loss rate, these continuity 
equations state that the  relative mass flows of the constituents correspond to their 
relative mass fractions in the stellar atmosphere. Only if this is true can the surface 
composition of the star be expected to remain unchanged.

In the present paper, metals are considered as trace elements so that 
$\rho_{2} \ll \rho_{1}$. As it is $\rho_{\rm e} \ll \rho_{1}$, 
it follows that $v_{1} \approx v$ from Eq. (27).
The velocity $v_{2}$ 
of the mean metal must be higher than $v_{1}$ for that momentum is given 
from the metals to hydrogen and helium. Thus, with $v_{2} = v$,  Eq. (39) 
yields an upper limit for the mass density $\rho_{2}$. With this value and 
with $G \left ( x \right ) = G_{\rm max}$, from Eq. (30) an upper limit 
for the mean collisional acceleration on hydrogen and helium can be derived:   
\begin{equation}
g_{\rm coll}^{\left ( \rm max \right )} = \frac {\zeta_{2} \dot M}{4 \pi r^{2} v}
                                           k_{12} G_{\rm max}  \, .
\end{equation}
For each wind model obtained from a one-component description and with 
$v_{1} \approx v$ and $\rho_{1} \approx \rho$ the lefthand 
side of the  momentum equation for hydrogen and helium 
(Eq. 24) is known at each grid point. Then the constituents may be coupled 
only if
\begin{equation}
g_{\rm coll}^{\left ( \rm max \right )} \geq v \frac {dv}{dr} 
+ \frac {1}{\rho} \frac {dp_{1}}{dr} + \frac {G M_{*}}{r^{2}} 
- \frac {1}{m_{1}} Z_{1} e E - g_{\rm rad}^{\left ( 1 \right )}   \, .
\end{equation}
If this condition is not fulfiled eyerywhere in the wind, then the wind 
solution is incompatible with the assumption that the wind is chemically 
homogeneous. In this case clearly a multicomponent description would be 
required. The mean radiative acceleration $g_{\rm rad}^{\left ( 1 \right )}$
on hydrogen and helium is obtained from own calculations with assumptions 
as described in Sect. 2.4.2. The results have shown that its effect is 
negligibly small. In Sects. 4 and 5, mass-loss rates are predicted as a 
function of surface gravity. If $g_{\rm rad}^{\left ( 1 \right )}$ were 
neglected in criterion (41), the maximal surface gravity up to which 
coupled winds can exist would be lower by about 0.05 dex only in $\log g$. 
%++++++++++++++++++++++++++++++++++++++++++++++++++++++++++++++++++++++
\subsection{Comparison with criteria from other authors}
%**************************************************************************
In condition (41) the contribution of the gas pressure to the momentum 
equation is taken into account. In the supersonic region, where this 
contribution is small, criterion (41) reduces to the one derived
e.g. by Owocki \& Puls (\cite{owo02}). If in Eq. (26) all terms proportional to 
$\rho_{\rm e}$ are neglected, it follows that
\begin{equation}
E \approx - \frac {1}{e n_{\rm e}} \frac {d p_{\rm e}}{dr}  \, .
\end{equation}
With this approximation the radiative acceleration 
$g_{\rm rad}^{\left ( \rm e \right )}$  due to the light scattering on 
free electrons has been neglected. This is justified in  
stars with $\Gamma_{\rm e} \ll 1$ (see Sect. 2).
If the metals are trace elements, it is 
$n_{\rm e} \approx Z_{1} n_{1}$. With $v_{1} \approx v$, from 
Eq. (28), it follows that $v_{\rm e} \approx v$. Thus hydrogen and helium, 
as well as the free electrons, move approximately with the centre of mass 
velocity $v$. If approximation (42) is inserted into condition (41), then with 
$p \approx p_{1} + p_{\rm e}$ the latter can be written as
\begin{equation}
g_{\rm coll}^{\left ( \rm max \right )} \geq v \frac {dv}{dr}
+ \frac {G M_{*}}{r^{2}} + \frac {1}{\rho} \frac {dp}{dr}
- g_{\rm rad}^{\left ( 1 \right )}  \, .
\end{equation}
If the contributions of the gas pressure and of the radiative force 
on hydrogen and helium are neglected, this criterion is equivalent to the 
one in Owocki \& Puls (\cite{owo02}). This is a good approximation 
in the supersonic 
region, whereas the criterion used in the present paper may also be applied in 
the subsonic region, where the contribution of the gas pressure is essential.
If the acceleration term $v dv / dr$ is also neglected, it follows that
\begin{equation}
g_{\rm coll}^{\left ( \rm max \right )} \geq \frac {GM_{*}}{r^{2}}  \, .
\end{equation}
In the supersonic region, this is a necessary but not sufficient condition for 
the existence of a coupled wind. 
It only requires that hydrogen and helium do not decelerate. In wind models 
obtained from the improved version of the CAK theory,   
the acceleration term should be more important than the gravitational one 
(Gayley \cite{gay00}). In the present wind models for high gravity 
stars obtained from the original version of the CAK theory, both terms 
have a similar order of magnitude in the supersonic region. 
\begin{figure}
\resizebox{\hsize}{!}{\includegraphics{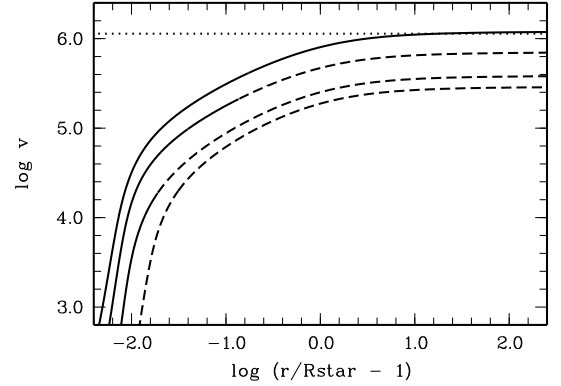}}
\\[1.0cm]
%\centering
%\includegraphics[angle=0,width=7.5cm]{kuf6.eps}
\caption{Solutions $v \left ( r \right )$ (in $\rm m \, \rm s^{-1}$) 
for $T_{\rm eff} = 40000 \, \rm K$, 
$\log g = 5.8$, $M_{*} = 0.5 M_{\odot}$, 
and (from top to the bottom) $Z/Z_{\odot} = 1$, $1/3$, $0.2$, and $0.1$. 
Dashed lines indicate decoupling of metals, the dotted line represents the 
surface escape velocity.}
\end{figure}

For a given mass-loss rate and metallicity, with 
Eq. (40) for $g_{\rm coll}^{\left ( \rm max \right )}$, a maximum velocity 
can be obtained up to which criterion (44) may be fulfiled:
\begin{equation}
v_{\rm max} = \frac {\zeta_{2} \dot M}{4 \pi G M_{*}} k_{12} G_{\rm max}  \, .
\end{equation}
This $v_{\rm max}$ is an upper limit of the velocity, up to which the constituents 
may be coupled. In accelerating winds decoupling should in fact occur at lower 
velocities.
Equation (45) corresponds to Eq. (22) of Owocki \& Puls (\cite{owo02}) 
for the case $w = 0$ in their notation.

This maximum velocity may be compared with the surface escape velocity:
\begin{equation}
v_{\rm esc} = \sqrt {\frac {2 G M_{*} \left ( 1 - \Gamma_{\rm e} \right )}
{R_{*}}}  \, .
\end{equation}
For $\Gamma_{\rm e} \ll 1$, with $R_{*} = \sqrt {G M_{*} g^{-1}}$, it follows that
\begin{equation}
v_{\rm esc} = \sqrt {2} \left ( G M_{*} g \right )^{\frac {1}{4}}  \, .
\end{equation}
With the definition
\begin{equation}
y = \frac {n_{2}}{n_{1}}
\end{equation}
and $\zeta_{2} \approx m_{2} m_{1}^{-1} y$, with Eq. (31) for $k_{12}$, 
the ratio $v_{\rm max} / v_{\rm esc}$ can be written as
\begin{equation}
\frac {v_{\rm max}}{v_{\rm esc}} = \dot M y 
\frac {Z_{1}^{2} Z_{2}^{2} e^{4} \ln \Lambda G_{\rm max}}{m_{1}^{2} k_{\rm B} T}
\frac {1}{\sqrt {2}} \left ( G M_{*} \right )^{- \frac {5}{4}} g^{- \frac {1}{4}} \, .
\end{equation}
For the wind models considered here, the Coulomb 
logarithm $\ln \Lambda$ varies between about $6.0$ at the wind base and $18.0$ 
in the outer regions of the thinnest winds. With the values assumed by  
Owocki \& Puls (\cite{owo02}), $\ln \Lambda = 20$, $Z_{1} = 1$ , 
$m_{1} = m_{\rm p}$, for a stellar mass $M_{*} = 0.5 M_{\odot}$ as considered 
here, it is
\begin{equation}
\frac {v_{\rm max}}{v_{\rm esc}} = 1.4*10^{20} \dot M y Z_{2}^{2} T^{-1} 
g^{- \frac {1}{4}}  \, ,
\end{equation}
where the mass-loss rate $\dot M$ is in $M_{\odot} / \rm yr $, and the surface 
gravity $g$ in $\rm cm \, \rm s^{-2}$. The temperature $T$ (in K) in the present 
paper is assumed to be equal to the effective temperature of the star.
Frictional heating could reduce $v_{\rm max}$ and thus favour 
decoupling, if the effect of a higher temperature is not compensated 
by a higher mean charge of the metals.
%*******************************************************************************
\subsection{Example: $T_{\rm eff} = 40000 \, \rm K$, $\log g = 5.8$}
%*******************************************************************************
%
%
For $T_{\rm eff} = 40000 \, \rm K$, $\log g = 5.8$, 
$M_{*} = 0.5 M_{\odot}$, and several metal abundances, the solutions 
$v \left ( r \right )$ are shown in Fig. 4. 
It is indicated in which parts of the solution condition (41) is fulfiled,  
and the maximum velocities $v_{\rm d}$ up to which this is the case are given 
in Table 1.
In these calculations the metals are represented by the CNO 
elements with force multipliers from own calculations as described in Sect. 2.4.2.  

The only case where 
the constituents may be coupled throughout the wind is for $Z / Z_{\odot} = 1$, for 
which a mass-loss rate $\log \dot M = -11.3$ is predicted. (This value agrees 
with the one obtained in Sect. 4 with the force multipliers from 
Kudritzki \cite{kud02}).  This mass-loss rate is lower 
by a factor of $\sim 1000$ than in the main sequence star $\tau \rm Sco$, for which 
Springmann \& Pauldrach (\cite{spr92}) used the values 
$\dot M = 5 * 10^{-9} M_{\odot} / \rm yr$, $M_{*} = 19.6 M_{\odot}$, 
$R_{*} = 5.5 R_{\odot}$ and $T_{\rm eff} = 33000 \, \rm K$ in their analysis of 
multicomponent effects, which corresponds to 
$\log g = 4.3$. As in the present example the stellar radius is only 
$R_{*} = 0.15 R_{\odot}$, the mass flux $\dot M / \left ( 4 \pi R_{*}^{2} \right )$ 
near the stellar surface has a similar order of magnitude to the one in $\tau \rm Sco$.
%+++++++++++++++++++++++++++++++++++++++++++++++++++++++++++++++++++++
\begin{table}
\begin{tabular}{l|llll}
$Z/ Z_{\odot}$ & $\log \dot M$ & $v_{\infty} / v_{\rm esc}$ & 
$v_{\rm d} / v_{\rm esc}$ (41) & $v_{\rm max} / v_{\rm esc} (50)$ \\
\hline
1.0 & -11.3 & 1.10 & -  & 5.64 \\
1/3 & -11.9 & 0.63 & 0.186 & 0.47 \\
1/5 & -12.9 & 0.35 & 0.017 & 0.028 \\
1/10 & -14.7 & 0.26 & 0.0004 & 0.0002 \\  
\end{tabular}
\caption{Predicted mass-loss rates (in $M_{\odot} / \rm yr$), terminal velocities, 
and velocities at which decoupling is expected  
according to criteria (41) and (50), respectively, for $T_{\rm eff} = 40000 \, \rm K$, 
$\log g = 5.8$, and various metal abundances. The surface escape velocity is 
$v_{\rm esc} = 1136 \, \rm km / \rm s$.} 
\end{table}
%+++++++++++++++++++++++++++++++++++++++++++++++++++++++++++++++++++++
Thus the wind densities are similar. Springmann \& Pauldrach 
(\cite{spr92}) and Krti\v cka \& Kub\'at (\cite{krt01b}) find that in $\tau \rm Sco$ 
the multicomponent effects are important due to their contribution to the 
energy equation. In the present example, the relative velocities between 
metals and the passive plasma are up to $7 \, \rm km / \rm s$. This may lead to  
frictional heating. Thus, as 
in $\tau \rm Sco$, the present example is near the border of the runaway 
condition. Multicomponent effects may have some importance; however, the 
coupling should still be effective enough the passive plasma to 
be expelled from the star.
\begin{figure*}
\centering
% ursprueglich width = 8.0 cm
\includegraphics[angle=0,width=6.5cm]{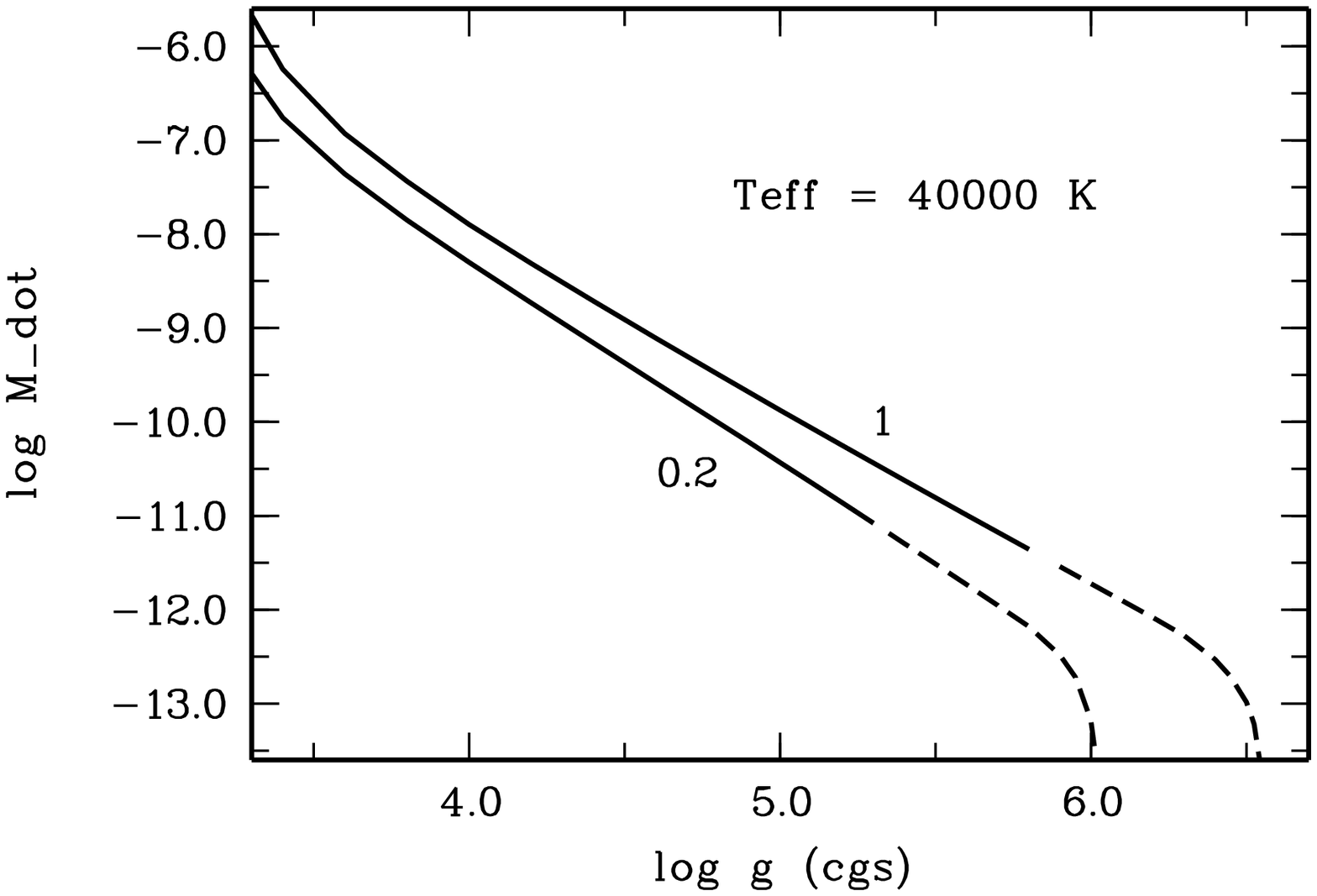}%
\hspace{0.5cm}%
\includegraphics[angle=0,width=6.5cm]{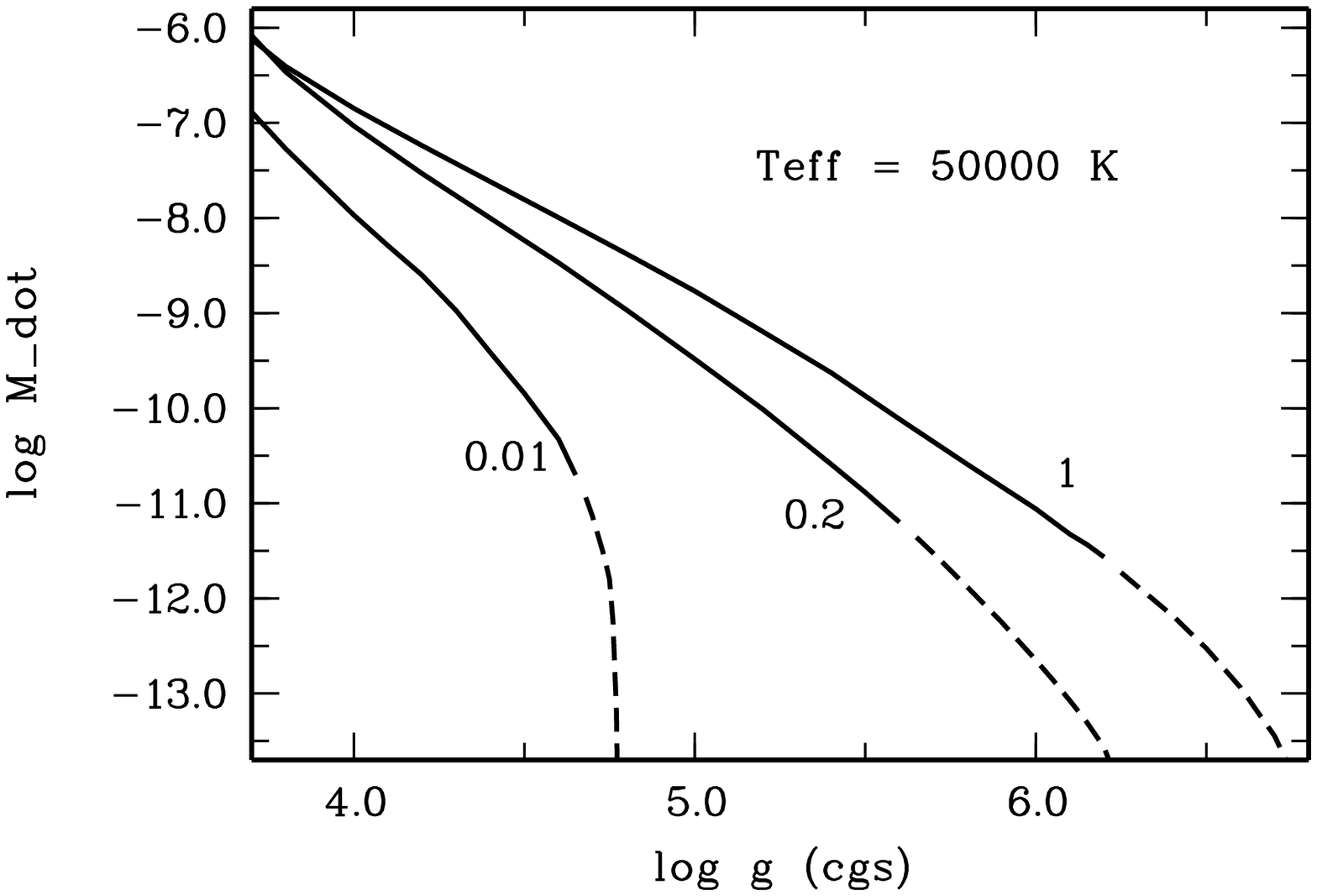}
\\[1.0cm]
\includegraphics[angle=0,width=6.5cm]{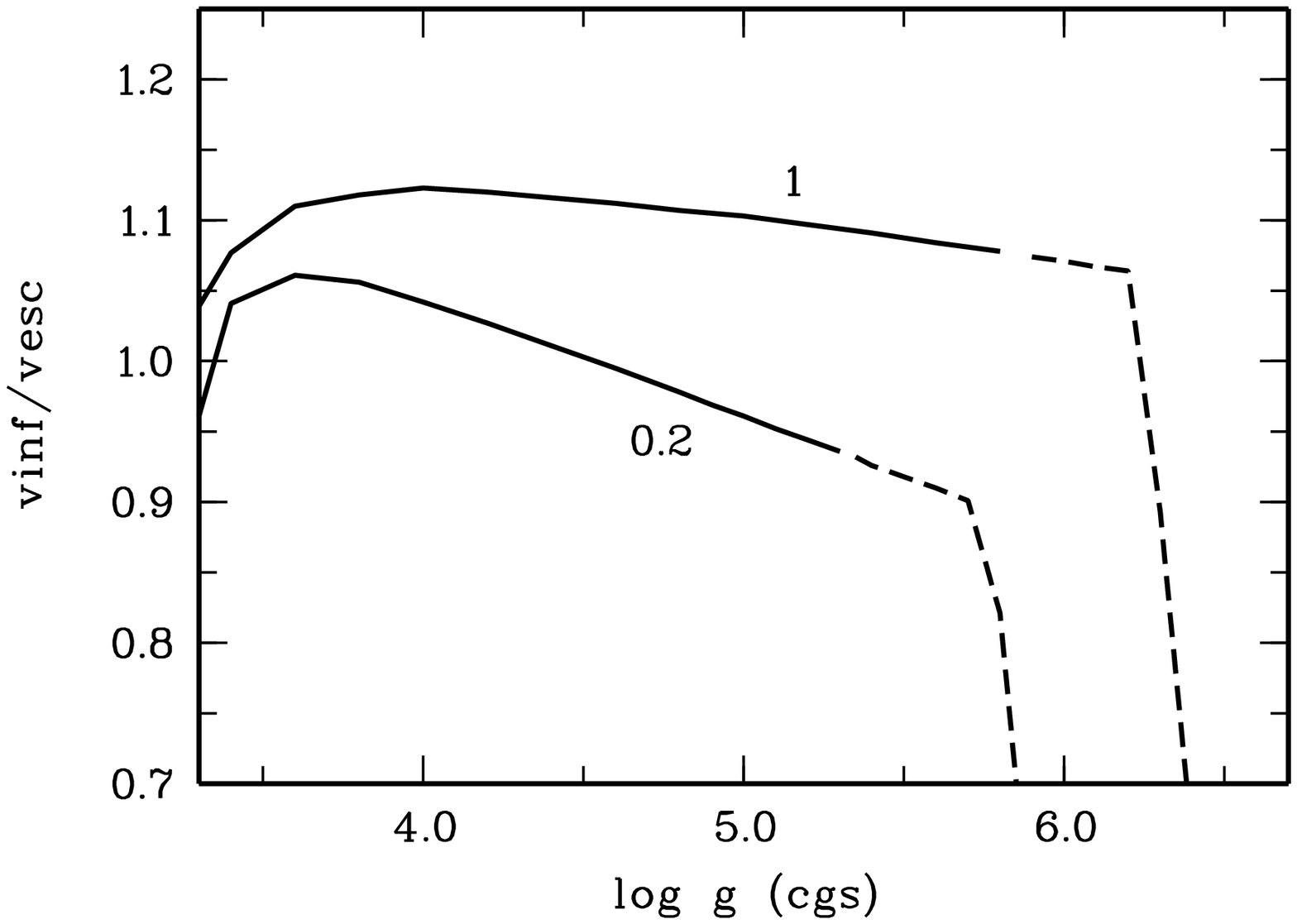}%
\hspace{0.5cm}%
\includegraphics[angle=0,width=6.5cm]{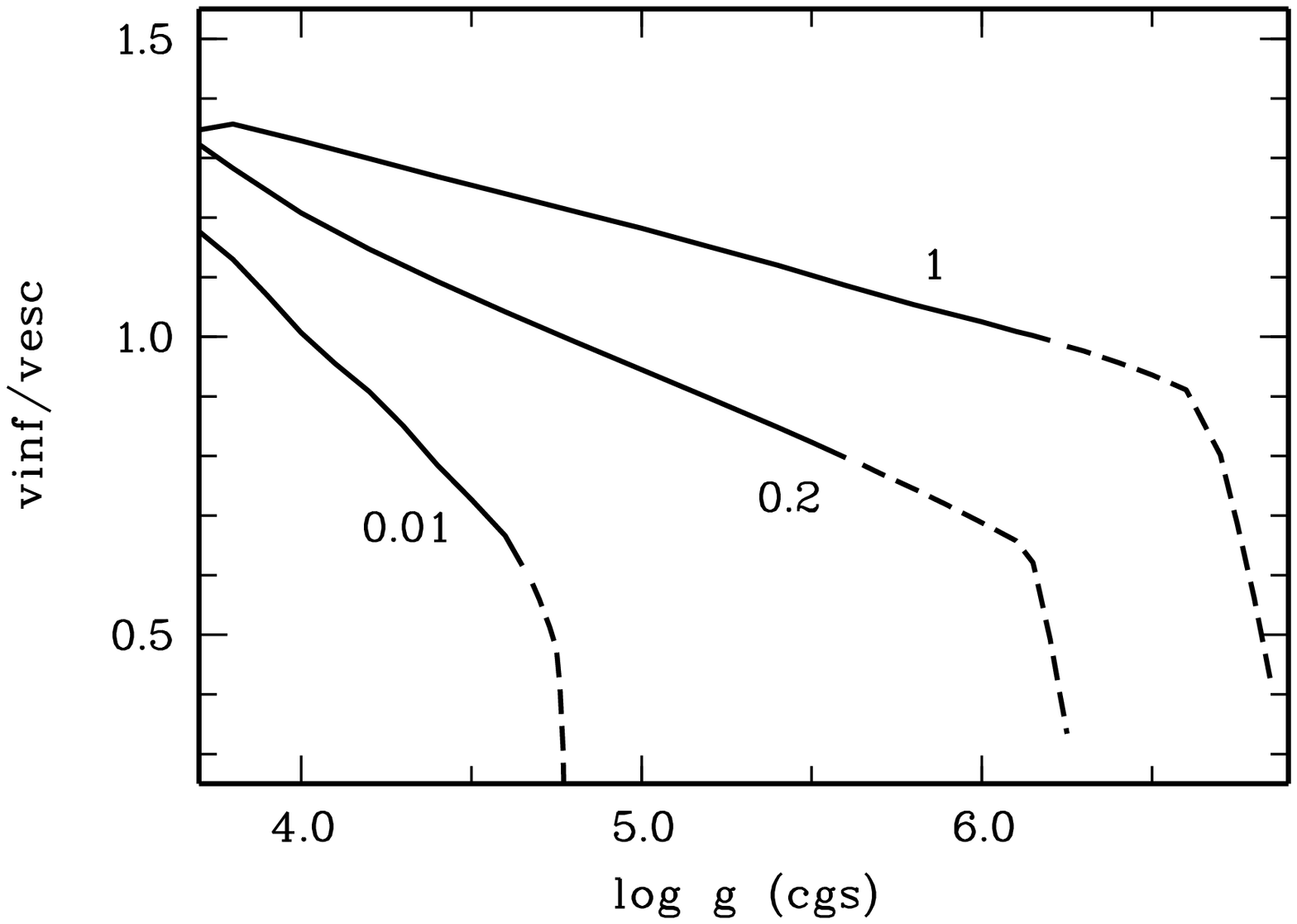}
\\[1.0cm]
\includegraphics[angle=0,width=6.5cm]{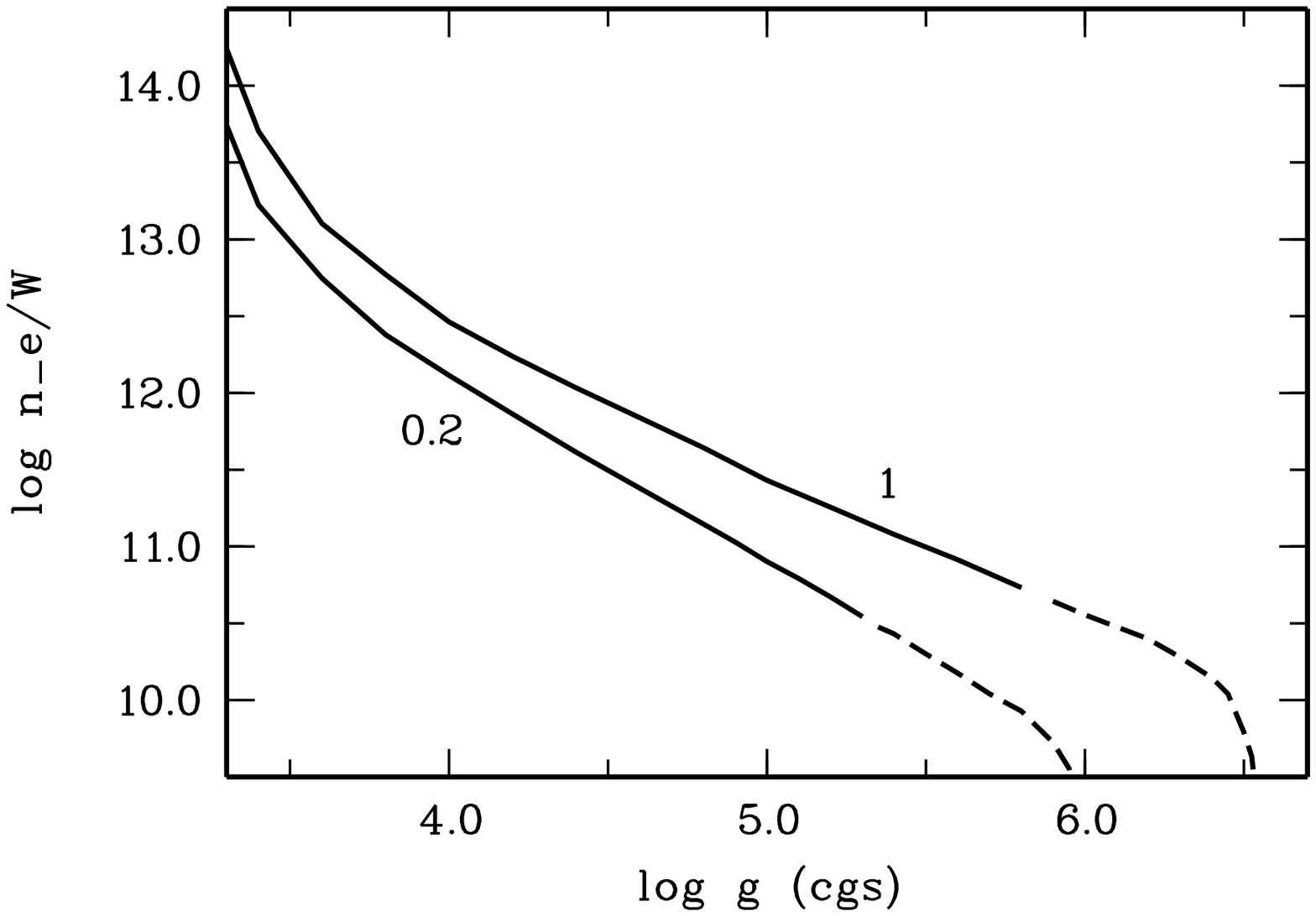}%
\hspace{0.5cm}%
\includegraphics[angle=0,width=6.5cm]{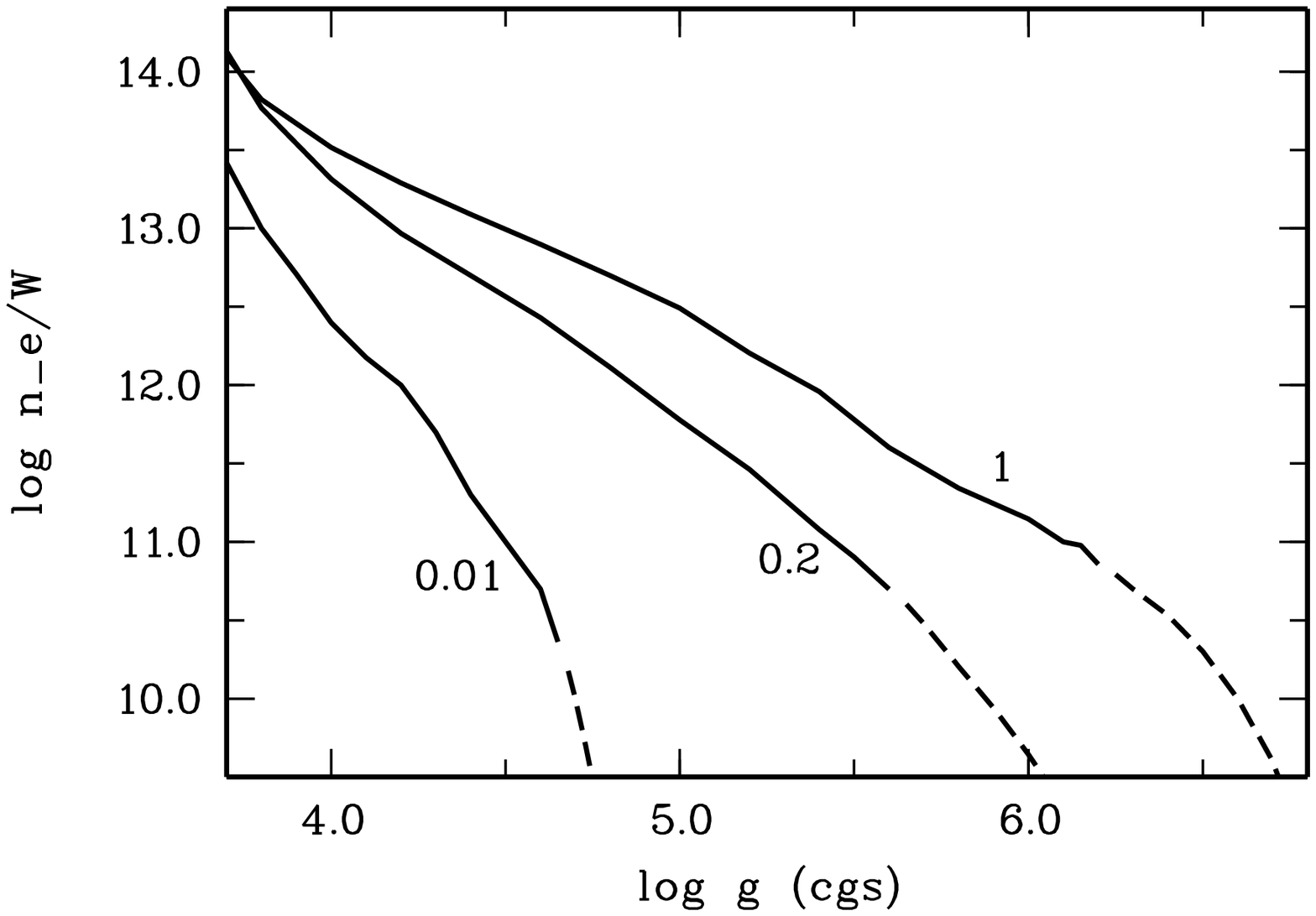}
\\[1.0cm]
\caption{Left: Predicted mass-loss rates (in $M_{\odot} / \rm yr$), 
terminal velocities, and the 
adopted values of $n_{\rm e} / W$ (in $\rm cm^{-3}$) as a function 
of the surface gravity 
for $T_{\rm eff} = 40000 \, \rm K$, $M_{*} = 0.5 M_{\odot}$, and 
$Z / Z_{\odot} = 1$ and $0.2$, respectively. Right: The same for 
$T_{\rm eff} = 50000 \, \rm K$ and $Z / Z_{\odot} = 1$, $0.2$, and $0.01$. 
The results for which 
decoupling of the metals is expected are represented by dashed lines.}
\end{figure*}

If the metal abundance is reduced by a factor of three, $\log \dot M = -11.9$ 
is predicted by the one-component wind model, which is lower by a factor of four 
as in the case 
with $Z /Z_{\odot} = 1$. Because of the reduced mass-loss rate and the additional 
reduction of the metal abundance, the density of metals in the wind effectively is 
lower by a factor of $12$. Therefore in this case decoupling is expected at a 
velocity $v_{\rm d} = 210 \, \rm km / \rm s$ and at a radius $r = 1.13 \, R_{*}$. 
As the decoupling occurs near the stellar surface at a velocity that is lower 
by about a factor of five than the surface escape velocity 
$v_{\rm esc} = 1136 \, \rm km / \rm s$, 
the passive plasma cannot be expelled from the star. 
For metal abundances reduced by more than a factor of five with 
predicted mass-loss rates $\dot M \la 10^{-13} M_{\odot} / \rm yr$, the decoupling is 
even expected in the subsonic region.

If the stellar parameters and the calcuated mass-loss rates are inserted into 
Eq. (50), with a solar number ratio of metals to hydrogen and helium 
$y_{\odot} \approx 10^{-3}$, it can be seen that this criterion for ion decoupling 
leads to similar conclusions. The values of $v_{\rm max} / \rm v_{\rm esc}$
are given in the last column of Table (1). It is $v_{\rm max} > v_{\infty}$ only 
for solar metallicity. For the cases with 
reduced metallicity, it is $v_{\rm max} < v_{\infty}$, although 
criterion (50) overestimates $v_{\rm max}$. Thus it follows that, at least for the 
cases with reduced metallicity, no coupled wind can exist. This could only be possible, 
if the terminal velocities were significantly lower than predicted from the present 
calculations.
 
Krti\v cka \& Kub\'at (\cite{krt00}) suggest that in some cases decoupling 
could  be 
avoided if the wind switches to a solution with an abrupt lowering of the 
velocity gradient. However, the physical relevance of such solutions has been 
questioned by 
Owocki \& Puls (\cite{owo02}) and Krti\v cka \& Kub\'at (\cite{krt02}), 
who argue that they are unstable. The time-dependent hydrodynamical 
calculations of Votruba et al. (\cite{vot07}) show that the ions 
decouple and are accelerated to high velocities, whereas the passive 
plasma decelerates.
  
Feldmeier \& Nikutta (\cite{feld06}) investigated the effect of radiative 
coupling between distant locations in winds with non-monotonic velocity 
laws and allowed for wind deceleration. They find that this could lead to 
a lowering of the terminal velocity, which may be below the surface escape 
velocity by a factor of the order two to three. However, the existence of 
chemically homogeneous winds with $\dot M \approx 10^{-13} M_{\odot} / \rm yr$, which 
have been assumed e.g. by Unglaub \& Bues (\cite{ub01}) to explain the helium abundances 
in sdB stars, would require the 
existence of winds with $v_{\infty} \approx 0.1 v_{\rm esc}$. 
Such a strong lowering of the terminal velocity seems to be unlikely.
As discussed in Sect. 6 from the various simplifications used in the present 
calculations, it should instead 
be expected that the terminal velocities are underestimated. Then the mass-loss rates 
required for the existence of a coupled wind would be higher. 
%
%******************************************************************************
\section{Results for $T_{\rm eff} = 40000$ and $50000 \, \rm K$} 
%******************************************************************************
%--------------------------------------------------------------------------
%
%
This section presents the mass-loss predictions as a function of the surface gravity
for $T_{\rm eff} = 40000 \, \rm K$ and $50000 \, \rm K$ and for 
$M_{*} = 0.5 M_{\odot}$ (see Fig. 5). The radiative acceleration is 
obtained with the force multipliers from Kudritzki (\cite{kud02}) for 
several metallicities. The adopted values of 
$n_{\rm e} / W$ are shown in the lower panels, which were obtained according to 
the iteration procedure described in Sect. 2.4.1. 
For each wind model, it was checked whether 
decoupling of metals from hydrogen and helium 
is expected according to criterion (41); the 
adopted mean charges of these constituents are given in Table 2. 
The results of the calculations for which the actual one 
component description of the wind may be sufficient, are represented by 
solid lines. 
\begin{table}
\begin{tabular}{cccc}
$T_{\rm eff}$ [K] & $\sigma_{\rm e}$ $[\rm cm^{2} \rm g^{-1}]$ & $Z_{1}$ & 
$Z_{2}$ \\
\hline
40000 & 0.33 & 1.08 & 2.95 \\
50000 & 0.34 & 1.09 & 3.51 \\
\hline
\end{tabular}
\caption{Adopted electron scattering opacities and mean charges $Z_{1}$ (hydrogen 
and helium) and $Z_{2}$ (metals).}
\end{table}
The surface gravities for which wind models were calculated range from 
values where the star would be near the Eddington limit 
($\Gamma_{\rm e} \approx 0.8$) to values for which no wind solution exists.

For $T_{\rm eff} = 40000 \, \rm K$, $Z/Z_{\odot} = 1$, coupled winds can exist 
up to $\log g = 5.8$ with a mass-loss rate of 
$\dot M = 4.4*10^{-12} M_{\odot} / \rm yr$. For $Z/Z_{\odot} = 0.2$, this
is only the case for $\log g \leq 5.3$. Due to the higher luminosities for 
$T_{\rm eff} = 50000 \, \rm K$, the mass-loss rates for given $\log g$ and 
$Z$ are higher than for $T_{\rm eff} = 40000 \, \rm K$. However, even for 
solar metallicity, chemically homogeneous winds can exist for 
$\log g \leq 6.2$ alone. Thus for DA white dwarfs in the considered range of 
effective temperatures, their existence can be excluded. For $\log g > 7.0$ 
no wind solution exists at all, because the radiative acceleration is too low 
even for wind optical depth parameters $t \rightarrow 0$. For 
$T_{\rm eff} = 50000 \, \rm K$, $Z/Z_{\odot} = 0.2$, and 
$Z/Z_{\odot} = 0.01$, we expect the existence of coupled winds for 
$\log g \leq 5.5$ and for $\log g \leq 4.6$, respectively. 
\begin{figure}
\centering
\includegraphics[angle=0,width=7.0cm]{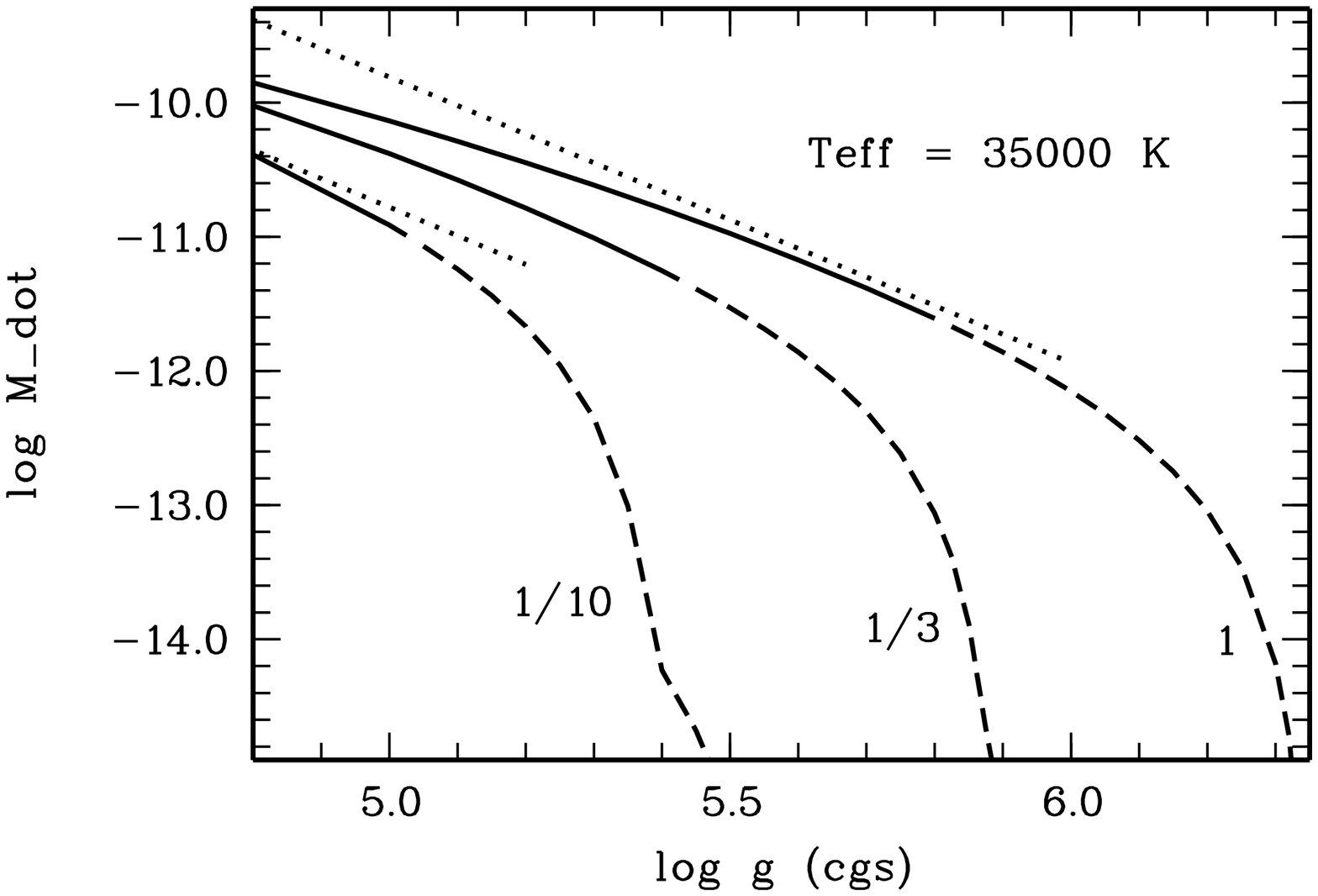}
\\[1.0cm]
\includegraphics[angle=0,width=7.0cm]{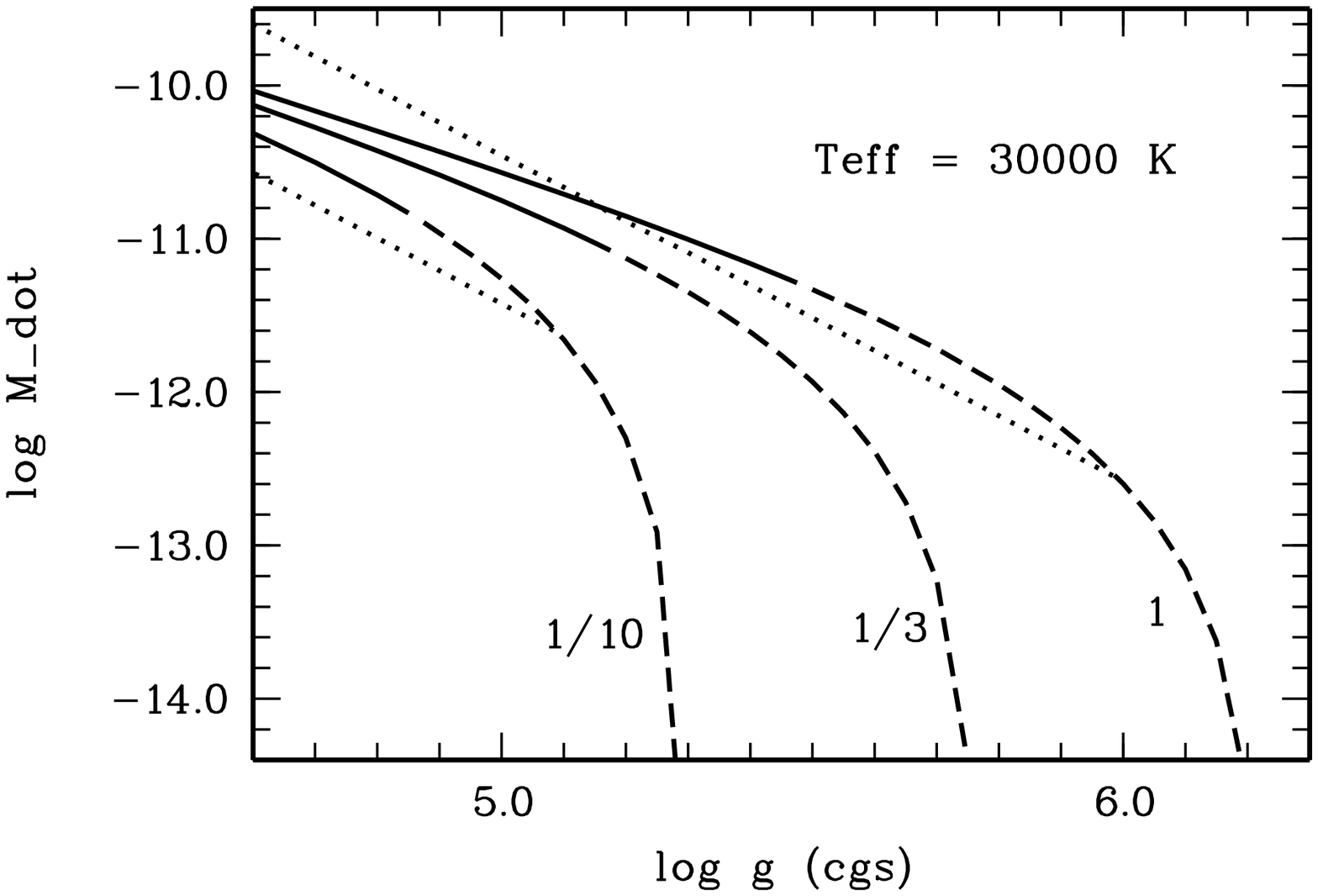}
\\[1.0cm]
\includegraphics[angle=0,width=7.0cm]{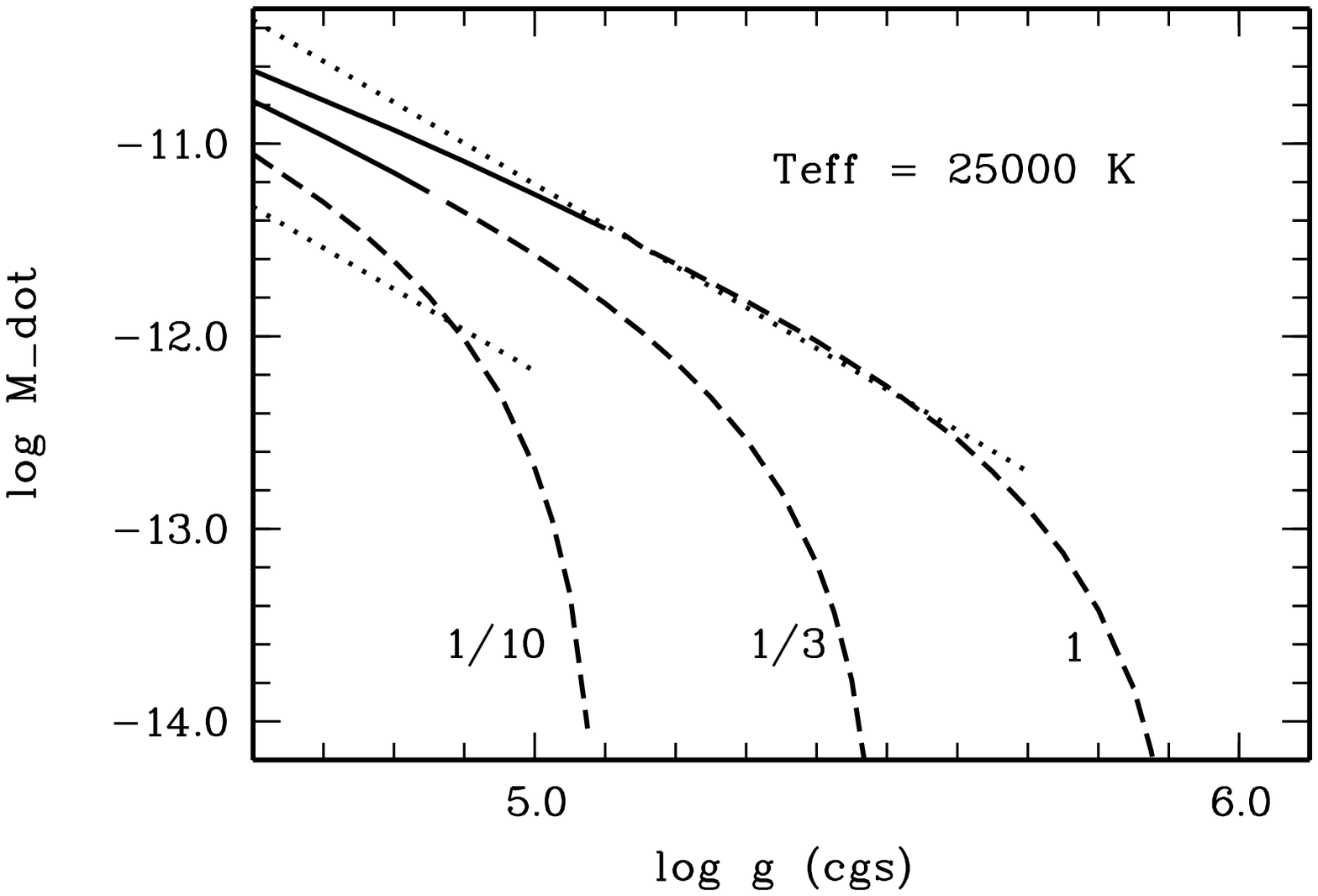}
\\[1.0cm]
\caption{Predicted mass-loss rates (in $M_{\odot} / \rm yr$) 
as a function of surface gravity for 
$T_{\rm eff} = 35000$, $30000$, $25000 \, \rm K$ and $Z/Z_{\odot} = 1$, $1/3$,
$0.1$ with $M_{*} = 0.5 M_{\odot}$. Dashed lines indicate 
decoupling of metals in the corresponding wind model. The upper and lower 
dotted lines in each figure represent the results of 
Vink \& Cassisi (\cite{vink02}) for $Z/Z_{\odot} = 1$ and $0.1$, 
respectively.} 
\end{figure}

The results show an increasing dependence of the mass-loss rates on 
metallicity to higher gravities.
For $T_{\rm eff} = 40000 \, \rm K$, $\log g = 4.0$, and $Z/Z_{\odot} = 1$, a 
mass-loss rate of $\dot M = 1.3*10^{-8} M_{\odot} / \rm yr$ is predicted. 
For $Z/Z_{\odot} = 0.2$  and the same stellar parameters, it is 
$\dot M = 5.0*10^{-9} M_{\odot} / \rm yr$. From a comparison of these two 
values, a dependence of the mass-loss rate on the metallicity according to 
$\dot M \sim Z^{0.58}$ may be suggested. For $\log g = 5.0$ and the same 
effective temperature the mass-loss rates for the two metallicities are 
$1.3*10^{-10}$ and $3.7*10^{-11} M_{\odot} / \rm yr$, respectively. From 
these values it would follow that $\dot M \sim Z^{0.80}$. This shows that, for 
weak winds, the Z-dependence cannot be represented by a power law of the 
form $\dot M \sim Z^{\epsilon}$ with constant exponent $\epsilon$. For the 
present example, $\epsilon$ would increase from $\epsilon \approx 0.6$ for 
the lowest gravities up to values  $\epsilon > 1$ for $\log g \ga 5.5$. 
This effect can still be seen more clearly for the case
$T_{\rm eff} = 50000 \, \rm K$ in Fig. 5, for which results for 
$Z/Z_{\odot} = 0.01$ are also shown. This breakdown of the 
usual scaling relations for sufficiently weak winds has already been 
found by Kudritzki (\cite{kud02}) in his calculations for stars with 
extremely low metallicities. 
 
From the middle panels of Fig. 5 it can be seen that the terminal 
velocities of the present wind models have a similar order of magnitude to 
the surface escape velocities. This is expected from the original version 
of the CAK theory, as used in the present paper. According to the 
improved version of the theory they should be larger (see Sect. 6.1).
With $M \left ( t \right ) = k t^{- \alpha}$, it follows from the original theory 
that $v_{\infty} = \sqrt { \frac {\alpha}{1 - \alpha}}$ (see e.g. 
Lamers \& Cassinelli \cite{lam99}). For $\alpha$ between about 
$0.5$ and $0.7$, ratios of $1.0 \la v_{\infty} / v_{\rm esc} \la 1.5$ 
should be expected. As for small wind optical depth parameters 
the $\log M {\left ( t \right )}$ - $\log t$ relation 
flattens and its slope $- \alpha$ finally approaches 
zero, the predicted ratio $v_{\infty} / v_{\rm esc}$ may be lower for weak winds, 
however.
The sudden decrease of $v_{\infty}/ v_{\rm esc}$ and in the mass-loss rates, 
which occur in some cases near the highest gravities for which a wind solution 
still exists (e.g. for $T_{\rm eff} = 40000 \rm K$, $Z/ Z_{\odot} = 1$ near
$\log g = 6.3$), are a consequence of the restriction of Kudritzki's 
force multipliers to a value $M_{\rm max}$ as described in Sect. 2.4.1. 
The results for lower gravities, however, are not affected by this 
restriction. 

The results of this section show that the existence of coupled winds requires 
mass-loss rates at least of the order $10^{-11} M_{\odot} / \rm yr$. 
For stars with 
$M_{*} = 0.5 M_{\odot}$, effective temperatures in the range 
$40000 \, \rm K \leq T_{\rm eff} \leq 50000 \, \rm K$, and solar metallicity, 
this is possible only for surface gravities $\log g \la 6.0$. Stars with stellar 
parameters within these ranges may be pre-white dwarfs, which are in an evolutionary 
stage between the asymptotic giant branch or the EHB and the white dwarf 
cooling sequence. Thus, if the metallicity is not too far below the solar value, 
the mass-loss rates should be high enough to prevent the effect of diffusion.
 
In no case can
chemically homogeneous winds with $\dot M \la 10^{-12} M_{\odot} / \rm yr$  
exist, for which the effect of diffusion would become effective according 
to e.g. Unglaub \& Bues (\cite{ub01}). That this is hardly possible 
can already be seen from Eq. (50). 
If e.g. for $T_{\rm eff} = 40000 \rm K$, $\log g = 6.0$ 
and $y = 10^{-3}$ (which approximately corresponds to solar metallicty) a 
mass-loss rate $\dot M = 10^{-12} M_{\odot} / \rm yr$ is inserted; 
then with $Z_{2} \approx 3.0$, it follows that $v_{\rm max} / v_{\rm esc} = 1$. 
Thus in winds with 
terminal velocities $v_{\infty} \ga v_{\rm esc}$, which should be expected 
at least according to the improved version of the CAK theory (see Sect. 6), it is 
$v_{\rm max} < v_{\infty}$, so the constituents cannot be coupled 
throughout the wind.  
% 
%*********************************************************************************
%************************************************************************************
\section{Results for sdB stars}
%************************************************************************************
%++++++++++++++++++++++++++++++++++++++++++++++++++++++++++++++++++++++++++++++++++++
\begin{figure}
\centering
\includegraphics[width=7.0cm]{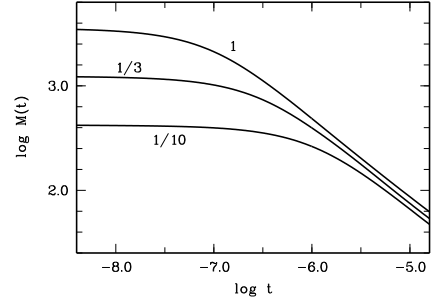}
\\[1.0cm]
\caption{Force multipliers according to our calculations as a function of the 
wind optical depth parameter for $T_{\rm eff} = 30000 \, \rm K$ and 
$Z/ Z_{\odot} = 1$, $1/3$, and $0.1$.}
\end{figure}

\begin{table}
\begin{tabular}{ccccc}
$T_{\rm eff}$ [K] & $Z / Z_{\odot}$ & $\sigma_{\rm e}$ $[\rm cm^{2} \rm g^{-1}]$
& $Z_{1}$ & $Z_{2}$  \\
\hline
35000 & 1 & 0.32 & 1.03 & 2.60 \\
35000 & 1/3 & 0.32 & 1.03 & 2.58 \\
35000 & 1/10 & 0.32 & 1.03 & 2.58 \\
\hline
30000 & 1 & 0.31 & 1.00 & 2.15 \\
30000 & 1/3 & 0.31 & 1.00 & 2.15 \\
30000 & 1/10 & 0.31 & 1.00 & 2.15 \\
\hline
25000 & 1 & 0.31 & 1.00 & 1.98 \\
25000 & 1/3 & 0.31 & 1.00 & 1.98 \\
25000 & 1/10 & 0.31 & 1.00 & 1.98 \\
\hline   
\end{tabular}
\caption{Adopted 
electron scattering opacities $\sigma_{\rm e}$, and mean charges $Z_{1}$ of hydrogen and helium 
and $Z_{2}$ of the metals, respectively.} 
\end{table}
%++++++++++++++++++++++++++++++++++++++++++++++++++++++++++++++++++++++++++++++++++
%
%
In Fig. 6 for $T_{\rm eff} = 35000 \, \rm K$, $30000 \, \rm K$, $25000 \, \rm K$, 
$Z/Z_{\odot} = 1$, $1/3$, $0.1$, and for $M_{*} = 0.5 M_{\odot}$, the predicted 
mass-loss rates are shown as a function of the surface gravity.
These results have been obtained with force multipliers from our own 
calculations (see Sect. 2.4.2). The adopted values for the electron scattering 
opacity and for the mean charges of hydrogen and helium and for the metals 
are given in Table 3. 

For the typical range of surface gravities of sdB stars 
($5.0 \la \log g \la 6.0$), the predicted mass loss rates are between about 
$10^{-10} M_{\odot} / \rm yr$ and zero. The results strongly depend on 
surface gravity and metallicity. At least for such cases with  
$\dot M \la 3*10^{-12} M_{\odot} / \rm yr$,  decoupling of the 
metals from the passive plasma is expected.
Thus the mass-loss rates may be 
high enough to prevent decoupling, but only in the most luminous sdB stars. 
For the more compact sdB stars from the one-component 
description of the wind mass-loss rates between about 
$10^{-12} M_{\odot} / \rm yr$ and zero are predicted. 
However, the constituents cannot be coupled in such weak winds.

Vink \& Cassisi (\cite{vink02}) derived mass-loss rates from the requirement of a 
global momentum balance with the assumption of a $\beta$-type velocity law with 
$v_{\infty} = v_{\rm esc}$. They calculated the radiative acceleration by means 
of a Monte Carlo simulation including about $10^{5}$ of the strongest lines of the 
elements H-Zn with NLTE occupation numbers for the most important elements.
As can be seen from Fig. 6, especially for $Z/Z_{\odot} = 1$ and for 
$5.0 \la \log g \la 6.0$, their results are in good agreement with the present 
ones. This may appear surprising, because Vink \& Cassisis's calculations of the 
radiative force are clearly more sophisticated. This point 
will be discussed in more detail in Sect. 5.1 for the case 
$T_{\rm eff} = 30000 \, \rm K$. 

The steep decrease in the mass-loss rates 
for $Z/Z_{\odot} = 0.1$ and $\log g \ga 5.0$ does not contradict  
Vink \& Cassisi's predictions, as it may seem if their results for more luminous 
stars were extrapolated to higher gravities. Calculations of Vink (priv. comm.) 
for cases with $\log g = 5.5$ and $Z/Z_{\odot} = 0.1$ have shown that no mass 
loss rate exists for which a global momentum balance can be fulfiled. Thus, 
if multicomponent effects are neglected, the mass-loss rate must approach 
zero for $\log g < 5.5$.
%+++++++++++++++++++++++++++++++++++++++++++++++++++++++++++++++++++++++++++++
\subsection{Detailed discussion for $T_{\rm eff} = 30000 \, \rm K$}
In Fig. 7 for $T_{\rm eff} = 30000 \, \rm K$ and $Z / Z_{\odot} = 1$, $1/3$, and 
$0.1$, the force multipliers according to own calculations are plotted as a 
function of the wind optical depth parameter. The mass-loss rates shown in the 
middle panel of Fig. 6 were calculated with these force multipliers.
The corresponding critical wind optical depth parameters (see Sect. 2.1)
are shown in Fig. 8. It can be seen that, for the considered surface gravities 
$4.6 \leq \log g \leq 6.2$, they are in the range $-5.5 > \log t_{\rm c}
\ga -8.0$. As can be seen from Fig. 7, the slope $- \alpha$ of the 
$\log M \left ( t \right )$ - $\log t$ relation is not constant in this range 
of wind optical depth parameters and $\alpha$ decreases to lower values of 
$t$. Thus, in a scaling law according to $\dot M \sim L^{\frac {1}{\alpha}}$ 
(Puls et al. \cite{puls00}), the exponent varies. An increasing 
dependence of the mass-loss rate on the luminosity to thinner winds is expected. 
In the present calculations for fixed $T_{\rm eff}$ and $M_{*}$, this explains why 
the dependence of the mass-loss rate on the surface gravity increases to 
higher values of $\log g$, as can be seen from the results in Fig. 6. 
\begin{figure}
\centering
\includegraphics[width=7.0cm]{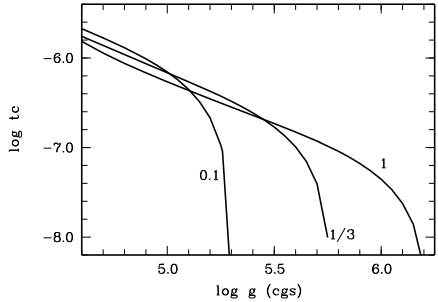}
\\[1.0cm]
\caption{Critical wind optical depth parameters as a function of the surface 
gravity for $T_{\rm eff} = 30000 \, \rm K$ and $Z/Z_{\odot} = 1$, $1/3$ and 
$0.1$.}
\end{figure}

In Fig. 9 for $T_{\rm eff} = 30000 \, \rm K$ and $Z/Z_{\odot} = 1$ and $0.1$, the 
contributions of the various elements taken into account in the calculations 
presented in this section to the force multiplier are plotted as a 
function of the wind optical depth parameter. It can be seen that in all cases
the major contribution is due to carbon 
for low values of $t$, which are relevant in the present calculations. 
From the present LTE assumption, it 
follows that $58$ \% of all carbon is C III and $42$ \% is C IV. The 
contribution of other ionization states of carbon is negligible.

In Fig. 10 for $Z/Z_{\odot} = 1$, the contributions of the 
lines C III $\lambda 977.0 \AA$, $\lambda 1175.7 \AA$ and of 
C IV $\lambda 1549.1 \AA$ to the force multiplier are shown as 
a function of $t$.
At $\log t = -6.0$, about $50$ \% of the total force multiplier is due 
to these three lines alone. For $t \rightarrow 0$ this contribution 
increases to $77$ \%. The most important one of these three lines, 
especially for $t \rightarrow 0$, is CIII $\lambda 977 \AA$.

Now let us discuss the special example $T_{\rm eff} = 30000 \, \rm K$, $\log g = 5.5$.
A coupled wind with a mass-loss rate 
$\dot M = 4.7 * 10^{-12} M_{\odot} / \rm yr$ is predicted.
The critical wind optical depth parameter is $\log t_{\rm c} = -6.7$. 
Here the contribution of these three carbon lines to 
the total force multiplier is $67$ \%. If all elements other than carbon 
were neglected, the derived mass-loss rate would be lower only by a factor of 
two ($\dot M = 2.5*10^{-12} M_{\odot} / \rm yr$). The effect of uncertainties 
in the CIII / CIV ionization equilibrium have a similar order of magnitude. 
If it were assumed that all carbon is in the ground state of C III, so that 
$\lambda 977 \AA$ is the only line to significantly contribute to the 
radiative force, this again would lead to $\dot M = 2.5 *10^{-12} 
M_{\odot} / \rm yr$. If on the other hand it is assumed that all carbon is  
in the ground state of C IV, this would reduce the mass 
loss rate by a factor of $3.6$ to $\dot M = 1.3 * 10^{-12} M_{\odot} / \rm yr$.
These results may explain the agreement between the present predictions 
and Vink \& Cassisi's within a factor of about two. As the major 
contribution to the radiative force is due to a few lines of carbon, the 
number of other lines and elements taken into account is not essential.
Secondly, the predicted mass-loss rates on the other hand always have the same order of 
magnitude, independent of the C III / CIV ionization equilibrium. 
If the ionization equilibrium changes, then the next ionization state takes 
over the contribution to the radiative force. 
\begin{figure}
\centering
\includegraphics[width=7.0cm]{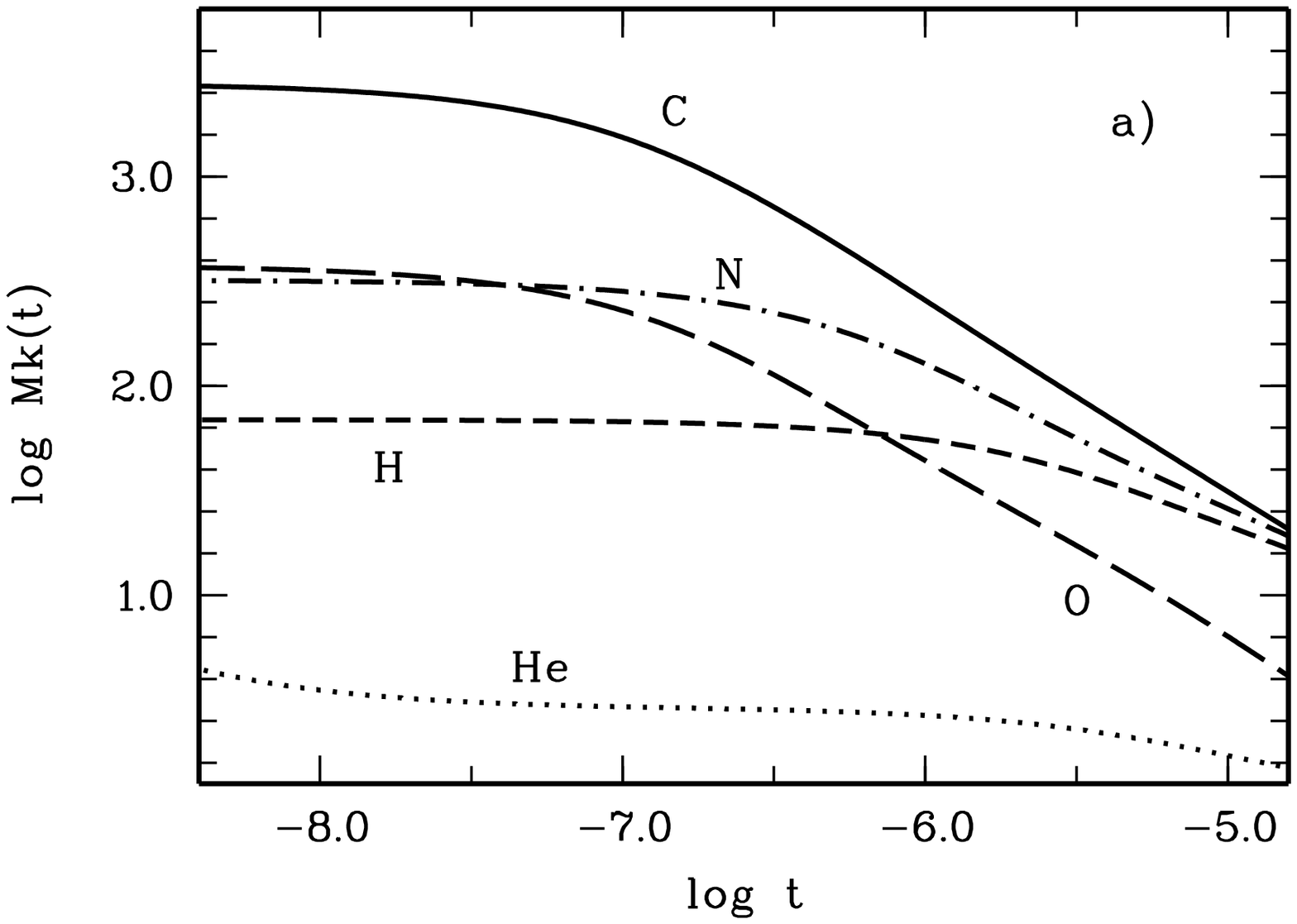}
\\[1.0cm]
\includegraphics[width=7.0cm]{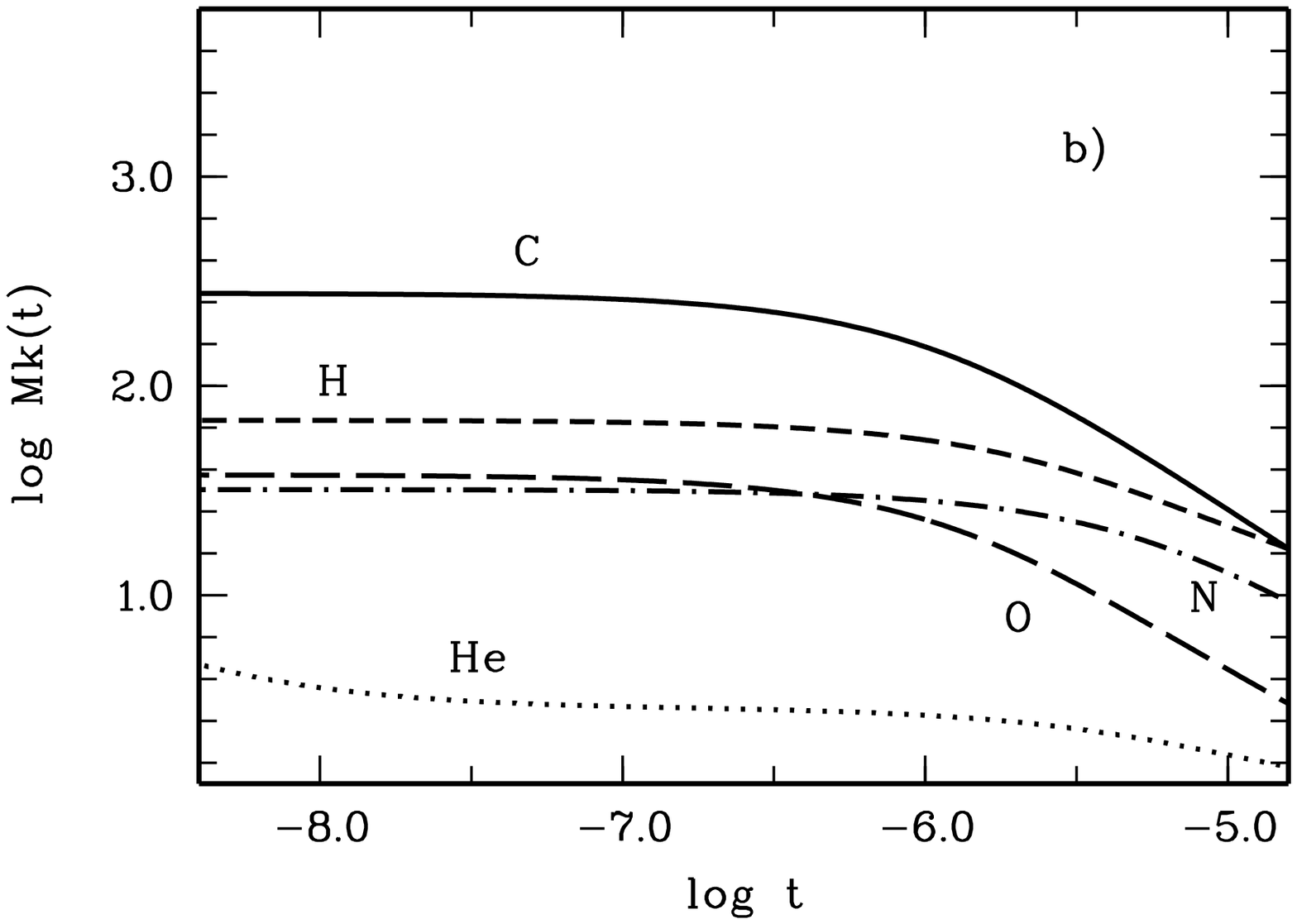}
\\[1.0cm]
\caption{Force multipliers due to the elements H, He, C, N, and O for 
$T_{\rm eff} = 30000 \, \rm K$ and for a) solar abundances, b) 
abundances of C, N, and O reduced by a factor of ten.}
\end{figure}

If the metal abundances are reduced by a factor of ten, then the major contribution 
to the force multiplier is still due to carbon (Fig. 9b). According to the present 
LTE calculations, the second largest contribution is due to hydrogen. 
For a wind optical depth parameter $\log t = -6.0$, about $71$ \% of this contribution 
is due to the line $\rm L \alpha$. Thus it essentially depends on the number 
density of the particles that are in the ground state of HI. The results shown in Fig. 9
were obtained with a number ratio of these particles to all hydrogen particles 
HI(n=1)/(HI+HII) $= 7.5*10^{-6}$. 
If the HI/HII ionization equilibrium is shifted to HII, then, in contrast to the 
CIII/CIV ionization equilibrium, the next ionization state cannot take over.
Thus a strong dependence on the ionization equilibrium is expected and the 
present results for hydrogen are questionable, because NLTE effects have not been 
taken into account. As the major 
contribution to the force multiplier for $Z/Z_{\odot} = 0.1$ is still due to carbon, however, 
the predicted mass-loss rates still agree approximately with the ones 
of Vink \& Cassisi (\cite{vink02}). A possible overestimate of the relative contribution 
of hydrogen to the force multiplier may be the reason that, for the cases with $\log g \la 5.0$  
shown in Fig. 6, the dependence of the mass-loss rates on the metallicity 
is somewhat less than in Vink \& Cassisi's results. 

For $T_{\rm eff} = 25000 \, \rm K$ the situation is similar to 
$T_{\rm eff} = 30000 \, \rm K$. The major contribution to the radiative 
force is due to carbon and the line C III $\lambda 977 \AA$. For 
$T_{\rm eff} = 35000 \, \rm K$ and wind optical depth parameters 
$\log t \la -6.0$ according to the present results, carbon is still 
the most important element. However, the contributions of N and O to 
the force multiplier are only slightly lower, by less than a factor of two.
\begin{figure}
\centering
\includegraphics[width=7.0cm]{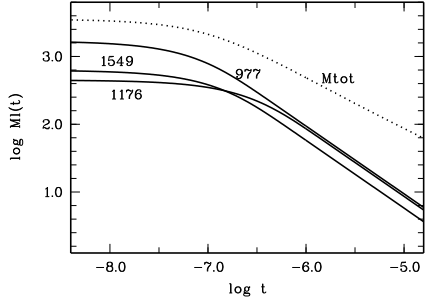}
\\[1.0cm]
\caption{Force multipliers due to the lines CIII $\lambda 977.0$ and 
$\lambda 1175.7 \AA$ and CIV $\lambda 1549.1 \AA$ for $T_{\rm eff} 
= 30000 \, \rm K$ and solar abundances. The dotted line represents the 
total force multiplier.}
\end{figure}
%++++++++++++++++++++++++++++++++++++++++++++++++++++++++++++++++++++++++++
\section{Discussion}
%++++++++++++++++++++++++++++++++++++++++++++++++++++++++++++++++++++++++++
In Sects. 6.1 and 6.2 we discuss the effects of the omission of the finite disk correction 
factor, of neglecting changes in ionization in the wind, and of 
the shadowing of the flux by the photospheric lines.  
Possible consequences of the results for the surface composition 
of the various types of chemically peculiar stars are then discussed in 
Sects. 6.3 and 6.4.
%+++++++++++++++++++++++++++++++++++++++++++++++++++++++++++++++++++++++
\subsection{Finite disk correction and changes in ionization}
The combined effect of the finite disk correction factor and changes in 
ionization may be estimated by comparing results from the 
present calculations with the theoretical predictions from other authors, who used 
the improved version of the CAK theory, which takes both effects into account. 
To eliminate the uncertainties of the force multipliers 
from our calculations, the values according to Kudritzki (\cite{kud02}) 
are used for this purpose. The inclusion of the finite disk correction alone 
should lead to lower mass-loss rates and higher terminal velocities 
by factors of 1.5 to four (Lamers \& Cassinelli \cite{lam99}). 

Kudritzki (\cite{kud02}) calculated mass-loss rates for massive and 
luminous stars. Two examples of stellar parameters for the case 
$T_{\rm eff} = 50000 \, \rm K$ are given in Table 4.
\begin{table}
\begin{tabular}{cccclc}
$\log L/L_{\odot}$ &  $\log g$ & $R_{*}$ & 
$M_{*}$ & $Z/Z_{\odot}$ & $\log n_{\rm e} / W$ \\
    & (cgs) & $[R_{\odot}]$ & $[M_{\odot}]$ & $[Z_{\odot}]$ & 
$[\rm cm^{-3}]$ \\
\hline 
7.03 & 3.63 & 43.76 & 298 & 1.0    & 14.11 \\
     &      &       &     & 0.2    & 14.15 \\
     &      &       &     & 0.01   & 13.42 \\
     &      &       &     & 0.001  & 12.69 \\
     &      &       &     & 0.0001 & 11.89 \\
6.42 & 3.85 & 21.71 & 122 & 1.0    & 13.54 \\
     &      &       &     & 0.2    & 13.36 \\
     &      &       &     & 0.01   & 12.54 \\
     &      &       &     & 0.001  & 11.61 \\
\hline 
\end{tabular}
\caption{Stellar parameters and adopted values of 
$n_{\rm e} / W$ for the comparison with the mass-loss 
predictions from Kudritzki (\cite{kud02}), with 
$T_{\rm eff} = 50000 \, \rm K$ in all cases.}
\end{table}
Ideally, our results for these stellar parameters should agree with 
Kudritzki's ones. As the finite disk 
correction and changes in ionization have been neglected in the present 
calculations, however, the results will usually disagree.
This can be seen from the comparison of both results in Fig. 11. 
In the last column 
of Table 4 the values of $n_{\rm e} / W$ obtained iteratively as described 
in Sect. 2.4.1 are tabulated. In the present calculations 
these values are assumed to be constant throughout the wind.

From the comparison it can be seen that our mass-loss rates in all cases 
are higher than Kudritzki's. 
In the various cases the discrepancies vary by a factor 
of $1.5$ up to a factor of seven.  
The terminal velocities are lower than Kudritzki's in most cases. Only for 
extremely low metallicities, the predicted terminal velocities approximately agree. 

These results correspond to the expectations.  
The effect of ionization may be more or less strong, depending on which elements and 
which ionization states primarily contribute to the radiative force. In some 
cases the combined effect of finite disk correction and ionization may lead to 
greater discrepancies than would be expected from the finite disk correction alone.
In other cases, both simplifications may almost compensate 
each other, so that the results agree approximately with 
results from the improved version of the CAK theory.
\begin{figure}
\centering
\includegraphics[angle=0,width=8.0cm]{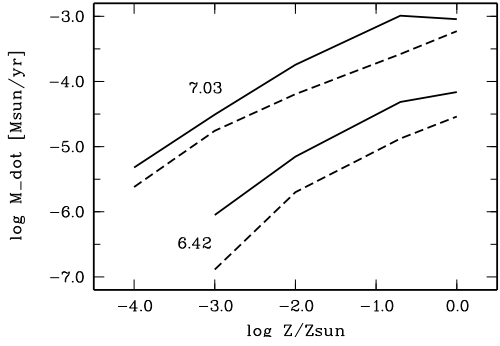}
\\[1.0cm]
\includegraphics[angle=0,width=8.0cm]{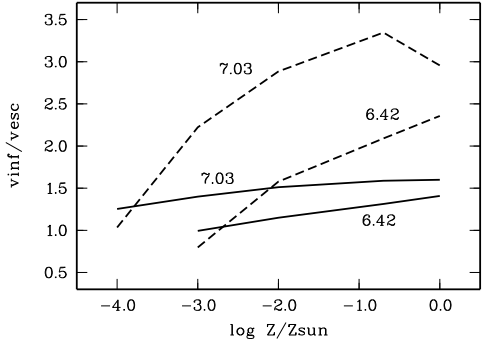}
\\[1.0cm]
\caption{Comparison of the predicted mass-loss rates
(top) and terminal velocities (bottom) as a function of metallicity  according to the 
present computational method (solid lines) with the results from 
Kudritzki (\cite{kud02}; dashed lines) for 
$\log L/L_{\odot} = 7.03$ and $6.42$, respectively, with $T_{\rm eff} = 50000 \, \rm K$.
In the present calculations, Kudritzki's 
force multipliers have been used as described in Sect. 2.4.1.} 
\end{figure}

In Table 5 mass-loss rates and terminal velocities obtained with the present 
computational method and force multipliers according to Kudritzki (\cite{kud02}) are 
compared with predictions from Pauldrach et al. (\cite{paul04}) and 
Pauldrach et al. (\cite{paul88}) for central stars of planetary nebulae 
with a solar composition. The adopted values of $n_{\rm e} / W$ are given in the 
last column.
\begin{table}
\begin{tabular}{ccccrc}
$T_{\rm eff}$ & $\log g$ & $M_{*}$ & $\log \dot M$ & $v_{\infty}$ & $\log 
n_{\rm e} / \rm W$ \\
$[\rm K]$ & (cgs) & $[M_{\odot}]$ & $[M_{\odot} / \rm yr]$ & $[\rm km / \rm s]$ 
& $[\rm cm^{-3}]$ \\
\hline
40000 & 3.70 & 0.410 & -7.28 &  296 & 12.9 \\
      &      &       & -7.74 &  420 &      \\
40000 & 3.80 & 1.110 & -7.10 &  431 & 12.7 \\
      &      &       & -7.21 &  850 &      \\
\hline
40000 & 3.35 & 1.000 & -5.73 &  194 & 13.8 \\
      &      &       & -5.82 &  400 &      \\
40000 & 3.69 & 0.644 & -7.06 &  332 & 12.9 \\
      &      &       & -7.42 &  800 &      \\
40000 & 4.45 & 0.546 & -8.78 &  580 & 12.0 \\
      &      &       & -8.85 & 1881 &      \\
50000 & 3.74 & 1.000 & -5.95 &  302 & 13.9 \\
      &      &       & -5.88 &  500 &      \\
50000 & 4.08 & 0.644 & -6.90 &  487 & 13.4 \\
      &      &       & -7.34 &  910 &      \\
50000 & 4.91 & 0.546 & -8.55 &  812 & 12.6 \\
      &      &       & -9.00 & 2583 &      \\
\hline  
\end{tabular}
\caption{Comparison of the mass-loss predictions for $Z / Z_{\odot} = 1$ 
with the results from 
Pauldrach et al. (\cite{paul04}) in the upper part of the table and 
Pauldrach et al. (\cite{paul88}) in the lower. For each set of stellar 
parameters the results from Pauldrach et al. are given in the lower line.}
\end{table}
In one case only are the mass-loss rates from our  
slightly lower (by a factor of 1.2) than the ones of Pauldrach et al. 
(\cite{paul88}). In all other cases, they are higher up to 
a factor of three. The terminal velocities 
in all cases are lower than the ones from Pauldrach et al. by factors between 
about $1.5$ and three. This again  agrees with the expected tendency of 
the predicted mass-loss rates of the present paper to be too large, 
whereas the terminal velocities are too low.
%++++++++++++++++++++++++++++++++++++++++++++++++++++++++++++++++++++++++++++++
\subsection{The effect of line shadowing}
%++++++++++++++++++++++++++++++++++++++++++++++++++++++++++++++++++++++++++++++
That the major contribution to the radiative force is due to a few strong lines 
for small wind optical depth parameters (see Sect. 5.1) has important consequences. 
In the present 
calculations the emergent monochromatic flux $F_{\nu}$ at the frequency of a line, 
from which the force multiplier is obtained (see Eq. (19)), corresponds to the continuum 
flux. In Fig. 12, for the line CIII $\lambda 977.0 \AA$ and for the case 
$T_{\rm eff} = 30000 \, \rm K$ , $\log g = 5.5$, and solar metal abundances, this flux is 
compared to the emergent flux from a LTE model atmosphere for sdB stars as
used for spectral analyses 
(Heber, priv. comm.). It can be seen that near the line centre this flux is lower by 
about a factor of $100$ than the flux that has been assumed in the mass-loss calculations.
Thus in the inner parts of the wind where the velocity and thus the Doppler shift 
is low, $F_{\nu}$ has clearly been overestimated. For larger Doppler shifts, both 
fluxes approach each other and approximately agree for
$\Delta \lambda \ga 3.5 \AA$. These Doppler shifts correspond to velocities 
$v \ga 1000 \, \rm km / \rm s$, which may be compared with the surface escape 
velocity $v_{\rm esc} = 955 \, \rm km / \rm s$ for this example with 
$\log g = 5.5$. Thus the flux assumed in the calculations is approximately correct
for the outer parts of the wind where $v \ga v_{\rm esc}$; however, it is clearly 
overestimated in the inner parts. 
  
According to the results of Babel (\cite{bab96}) this effect of ``line shadowing" 
should lead to lower mass-loss rates and higher terminal velocities in 
comparison to the usual theory that assumes a velocity-independent flux.
Then terminal velocities of the order $v_{\infty} \approx 5 v_{\rm esc}$ and
mass-loss rates that are lower by a factor of the order ten are not unusual.
Thus we expect that the present calculations overestimate the mass-loss rates.
The agreement with the results of Vink \& Cassisi (\cite{vink02}) does not  
contradict this, because the terminal velocity is a free parameter in their 
calculations. If a terminal velocity $v_{\infty} = 5 v_{\rm esc}$ instead 
of $v = v_{\rm esc}$ were assumed, then it follows from the dependence of 
$\dot M$ on the assumed value of $v_{\infty}$ derived in Vink et al. (\cite{vink00})
that the mass-loss rates may indeed be lower by a factor around ten. 
The consequence of lower mass-loss rates would be that decoupling of the 
metals occurs at lower gravities than predicted according to the 
present results.

Qualitatively, the inclusion of the effect of line shadowing should have similar consequences 
to including the finite disk correction (lower mass-loss rates, higher terminal velocities). 
Quantitatively, however, it seems to be much stronger for the weak winds discussed in the 
present paper. According to Owocki \& ud-Doula (\cite{owo04}) the finite disk correction 
reduces the radiative force in the inner parts of the wind only by a factor 
of about two, whereas the example shown in Fig. 12 shows that line shadowing may reduce 
the radiative force due to a strong line up to two orders of magnitude.
\begin{figure}
\centering
\includegraphics[width=7.0cm]{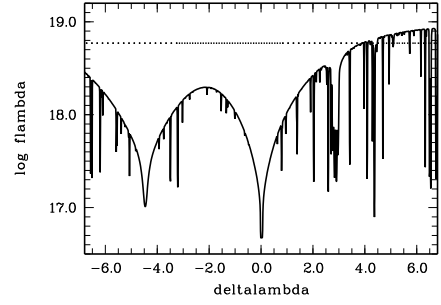}
\\[1.0cm]
\caption{Physical flux (in $\rm erg \, \rm cm^{-3} \rm s^{-1}$) near the 
line CIII $\lambda 977.0 \AA$ for 
$T_{\rm eff} = 30000 \, \rm K$, $\log g = 5.5$ and solar metal abundances 
as a function of $\Delta \lambda = \lambda - 977.0 \, \AA$. 
The solid line represents the flux from a model atmosphere (Heber, priv. comm.), 
the dotted line represents the flux used in the present calculations.}
\end{figure}

The results of Babel (\cite{bab96}) predict increasing 
terminal velocities for increasing surface gravities and otherwise fixed 
stellar parameters, whereas the present results shown in Sect. 4 predict 
a decreasing tendency for $v_{\infty}$. The effect of line shadowing probably also
changes the dependencies of the mass-loss rates and of the terminal velocities on 
the metallicity.
%++++++++++++++++++++++++++++++++++++++++++++++++++++++++++++++++++++++ 
\subsection{Consequences for the chemical composition}
The results presented in Sects. 4 and 5 have shown that, in hot white dwarfs and 
in the majority of sdB stars (see Sect. 7), no chemically homogeneous winds can exist, 
at least if the metallicity is not higher than solar. Thus, if any mass-loss exists, it 
should change the surface composition. The only exception may be that the decoupling 
of the metals from the passive plasma occurs at a radius, at which the velocity of 
hydrogen and helium already exceeds the local escape velocity. Then hydrogen and 
helium can still be expelled from the star. 
For the case where decoupling occurs before the 
local escape velocity has been reached, Porter \& Skouza (\cite{por99}) 
propose a periodic time-dependent scenario. The passive plasma decelerates after 
decoupling and eventually stalls. Thus a shell of gas is generated that 
reaccretes to the star. In the course of time, this scenario should lead to 
a depletion of metals near the stellar surface.

Another possible scenario is a pure metallic wind with hydrostatic hydrogen and 
helium. For main sequence A stars, this has been discussed by Babel (\cite{bab95}).
If the stellar atmosphere is approximately in a stationary state, then 
the metals must have an outward movement. This leads to an outward frictional 
force on hydrogen and helium. If these elements are in hydrostatic equilibrium 
($v_{1} = 0$) and if we demand that their partial pressure decreases in an outward 
direction ($dp_{1} / dr < 0$), then it follows from the momentum equation (24) for hydrogen 
and helium that the outward frictional acceleration must be less 
than the gravitational one: 
\begin{equation}
g_{\rm coll}^{\left ( 1 \right )} < \frac {G M_{*}}{r^{2}} \, .
\end{equation}
Otherwise the outward frictional force on the passive plasma due to the 
outflowing metals would be so strong that it cannot be in hydrostatic 
equilibrium.
As the density in the stellar atmosphere is much higher than in the wind, 
it is a reasonable assumption that the drift velocities 
of the metals are small in comparison to the thermal velocity.  
Then the collisional acceleration on hydrogen and helium can be 
calculated from the linear approximation.
For $x \ll 1$ it is 
$G \left ( x \right ) \approx \frac {2}{3} \frac {1}{\sqrt {\pi}} x$, so 
that the momentum exchanged per unit volume and unit time via Coulomb 
collisions may be written as
\begin{equation}
|\Delta Q| = n_{1} n_{2} l_{12} |v_{2} - v_{1}| 
\end{equation}
with
\begin{equation}
l_{12} = \frac {16}{3} \sqrt {\pi} 
\frac {Z_{1}^{2} Z_{2}^{2} e^{4}}{m_{12} \alpha^{3}} \ln \Lambda  \, .
\end{equation}
The quantity $n_{1} n_{2} l_{12}$ corresponds to the resistance 
coefficient $K_{\rm s \rm t}$ derived by Burgers (\cite{bur69}; 
see his Eq. 24.14).
With Eqs. (29), (52), (53), with $v_{1} = 0$, and the equation of continuity 
for the mean metal, it 
follows for the collisional acceleration on hydrogen and helium that
\begin{equation}
g_{\rm coll}^{\left ( 1 \right )} = \frac {1}{m_{1} m_{2}} 
\frac {\dot M_{2}}{4 \pi r^{2}} l_{12}  \, .
\end{equation}
Then condition (51) can be written as
\begin{equation}
\dot M_{2} < 4 \pi G M_{*} \frac {m_{1} m_{2}}{l_{12}}  \, .
\end{equation}
If the mass-loss rate of the metals fulfils this condition, then the 
collisional acceleration on the passive plasma never can exceed gravity, 
independent of how high the velocity of the mean metal is (because the 
linear approximation leads to an upper limit of  
$g_{\rm coll}^{\left ( 1 \right )}$). 
With $M_{*} = 0.5 M_{\odot}$, mean charges $Z_{1} = 1.0$, $Z_{2} = 3.0$, 
mean masses $m_{1} = 1.3 m_{\rm p}$, $m_{2} = 14.6 m_{\rm p}$, 
and with $\ln \Lambda = 12.0$, it follows for a temperature 
$T = 35000 \rm K$ that
\begin{displaymath}
\dot M_{2} < 3 * 10^{-16} \frac {M_{\odot}}{\rm yr}  \, .
\end{displaymath}
Thus hydrogen and helium may be in hydrostatic 
equilibrium if the mass-loss rate of the metals is lower than about 
$10^{-16} M_{\odot} / \rm yr$.
On the other hand, we know from the previous results that the existence 
of a coupled wind requires a total mass-loss rate greater than at least 
$10^{-12} M_{\odot} / \rm yr$. For an approximately solar mass fraction of 
the metals of the order of $10^{-2}$, this means that a coupled wind requires 
a mass-loss rate of the metals alone $\dot M_{2} > 10^{-14} M_{\odot} / \rm yr$.
If, however, $\dot M_{2}$ were somewhere in between these limits, 
e.g. $\dot M_{2} \approx 10^{-15} M_{\odot} / \rm yr$, then neither of the two 
cases seems to be possible. This value is too low for the existence of a 
coupled wind and too high for the case with hydrostatic passive plasma. 
Then hydrogen and helium may either fall back onto the star or be expelled, 
depending on at which radius decoupling occurs. Which of the various scenarios is 
the appropriate one essentially depends on the flow of the metals. 
In the following we discuss 
the case with hydrostatic hydrogen and helium and with a pure metallic wind. 
If the metals are trace elements and the densities in the wind are sufficiently 
low, then the mass-loss rates of the various metals should be 
independent of each other.
 
If the mass-loss rate of a metal is greater than zero and if the atmosphere 
is approximately in a stationary state, the metal must have an outward 
velocity. If the effect of concentration gradients is weak, then the radiative 
force must not only balance gravitational 
settling, but also must compensate the inward frictional force on the 
metal due to collisions with the passive plasma. Due to saturation effects, 
the radiative force on a metal increases with decreasing abundance.
Consequently, the abundances of these metals with non-zero mass-loss rates 
tend to be lower than the abundances obtained from 
diffusion calculations that assume an 
equilibrium between the downward gravitational force and the upward radiative force.
Only for these metals, for which the mass-loss rate is sufficiently low, should the 
equilibrium abundances agree with the ones derived from spectral analyses.

Seaton (\cite{sea96}, \cite{sea99}) presented the results of time-dependent 
diffusion calculations for iron group elements in the stellar envelopes of 
HgMn stars and allowed for an outflow of these elements at the outer boundary. 
The abundances are predicted to vary with time.
It may in principle happen that two stars with similar stellar parameters 
and surface compositions have different internal compositions. 
In the absence of strong 
concentration gradients, a maximum possible value for the 
outward flow of a metal can be derived for each depth, if the radiative 
acceleration is known as a function of the abundance. This maximum of the 
flow is small in regions where the metal has a noble gas configuration and 
thus the radiative acceleration is low. In 
Seaton (\cite{sea99}), these regions are referred to as barriers. 
How the surface abundance of a metal changes with time should depend essentially  
on its mass-loss rate and on the location of these barriers in the stellar 
envelope.

The abundance patterns of the various types of chemically peculiar stars 
discussed in the present paper qualitatively agree with these 
expectations, although exceptions exist. For sdB stars, equilibrium abundances 
have been predicted and compared with measured ones e.g. by 
Bergeron et al. (\cite{ber88}), Charpinet et al. (\cite{char97}), 
Ohl et al. (\cite{ohl00}), and Behara \& Jeffery (\cite{ntb07}). 
For some elements (e.g. Fe and N) there is  generally 
good agreement, whereas for others (e.g. Si) the predicted values 
may be too high by several orders of magnitude. According to
Chayer et al. (\cite{chay06}), the measured abundances of Ge, Zr, and Pb  
are somewhere between the solar value and the predicted equilibrium 
abundances, which for these elements exceed the solar value by about 
two orders of magnitude. The wide spread of the measured abundances may be 
an indication of a time-dependent scenario. 

For hot hydrogen-rich white dwarfs, metal abundances have been measured or 
predicted e.g. by Barstow et al. (\cite{bar03}, \cite{bar05}), Chayer et al. 
(\cite{chay9a}, \cite{chay9b}), Good et al. (\cite{good05}), 
Schuh et al. (\cite{son05}), Dobbie et al. (\cite{dob05}), and 
Vennes et al. (\cite{ven05}, \cite{ven06}). For some elements 
(Fe and O), the agreement is 
close in many cases, for others (e.g. C, N, and Ni) the abundances predicted from 
equilibrium diffusion theory tend to be too large. An exception is 
silicon, for which the measured abundances are usually larger than 
the predicted ones. Thus in hot hydrogen-rich white dwarfs and in some 
helium-rich ones (Dreizler \cite{drei99}) the situation is in some respect 
similar to sdB stars. The predicted abundances tend to be larger 
than the measured ones.

Another example is found in the HgMn stars. The manganese abundances in the stars  
analysed by Jomaron et al. (\cite{jom99}) in all cases are lower than 
predicted from the equilibrium diffusion calculations of 
Alecian \& Michaud (\cite{alec81}) and greater than the solar values.
Nitrogen is typically depleted by at least a factor of $100$, which is more 
than predicted from diffusion theory (Roby et al. \cite{roby99}).
The abundances of neon scatter from slight deficiencies to underabundances 
by one order of magnitude or even more (Dworetsky \& Budaj \cite{dwo00}).
This may point to some time-dependent process.  
The results of Seaton (\cite{sea96}, \cite{sea99}) for various iron group 
elements show that the presence of an outflow at the outer boundary  
in general leads to better agreement. An exception to the expected 
tendency is mercury, for which abundances greater than predicted from 
equilibrium diffusion theory have been detected (Proffitt et al., 
\cite{prof99}). However, at least in some HgMn stars, elements 
seem to be distributed inhomogeneously over the surface (Hubrig et al. 
\cite{hub06}), which complicates the problem.
%+++++++++++++++++++++++++++++++++++++++++++++++++++++++++++++++++
\subsection{Selective winds or turbulence?}
%+++++++++++++++++++++++++++++++++++++++++++++++++++++++++++++++++
As explained in the preceeding section, in the presence of pure metallic 
winds the abundances of the metals tend to be lower than 
predicted from equilibrium diffusion calculations. This should be so for 
overabundant, as well as for deficient, metals. An alternative explanation 
of the discrepancies between measured and predicted abundances 
may be the presence of turbulence, which could stem from mixing processes 
like convection or stellar rotation. Then, however, the abundances 
of the deficient metals 
should be larger than predicted from equilibrium diffusion calculations 
(Vauclair et al. \cite{vvm78}), because turbulence tends to 
level out concentration gradients and reduces the effect of gravitational 
settling. Additional mixing due to turbulence outside of convection zones 
has been assumed by Richer et al. (\cite{rich00}) in their calculations for 
AmFm stars (which represent the continuation of the HgMn phenomenon at lower 
effective temperatures) to reduce the amplitude of the predicted abundance 
anomalies to a level that agrees with observational results. 

In hot hydrogen-rich white dwarfs and sdB 
stars, however, the abundances not only of enriched metals are lower than 
predicted, but also those of deficient metals. In particular, 
the effect of turbulence alone can hardly explain why elements that should 
be overabundant according to diffusion theory are in fact deficient, as has 
been found for argon in two of the hottest known DA white dwarfs 
(Werner et al. \cite{wer07}). Thus the scenario with 
selective winds seems to be the most likely one for these stars.

The only element that is generally less deficient than predicted from 
equilibrium diffusion calculations, is helium.  This is so for these hot 
white dwarfs, in which helium is detectable (Vennes et al. \cite{ven88}), 
sdB (Michaud et al. \cite{mic89}), and HgMn stars (Michaud et al. \cite{mic79}). 
According to the results of Krti\v cka (\cite{kkg06}) for peculiar B stars, 
this can hardly be explained within the framework of selective winds.
Thus the helium deficiencies should be due to gravitational 
settling. It is possible that completely undisturbed stellar atmospheres hardly exist 
and some turbulence is always present. This may explain that helium sinks 
more slowly than expected. However, it is still unclear what the 
origin of this turbulence is exactly.
%**************************************************************************
\section{Summary and conclusions}
In the $T_{\rm eff}$-$\log g$-diagram of Fig. 13 for 
$25000 \, \rm K \leq T_{\rm eff} \leq 40000 \, \rm K$ and for 
$Z/Z_{\odot} = 1$, $1/3$ and $0.1$, the lines are shown above 
which chemically homogeneous winds may exist according to the present results.
The samples of sdB stars analysed by Maxted et al. (\cite{max01}) and 
Lisker et al. (\cite{lisk05}) shown in the figure have been analysed for 
peculiar $\rm H_{\alpha}$ line profiles, which may be interpreted as wind 
signatures according to Heber et al. 
(\cite{heb03}) and Vink (\cite{vink04}). As can be seen in the figure, 
such signatures have only been detected 
in the most luminous sdB stars.
According to the calculations of stellar evolution by Dorman et al. 
(\cite{dor93}), these ones are in a post-EHB stage of evolution.
\begin{figure}
\resizebox{\hsize}{!}{\includegraphics{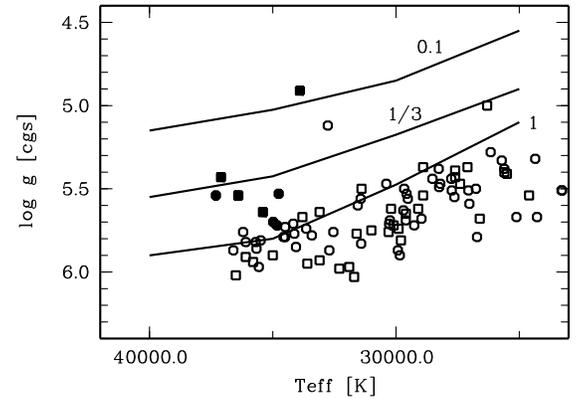}}
\\[1.0cm]
\caption{Lines in the $T_{\rm eff}$ - $\log g$ diagram above which 
chemically homogeneous winds may exist for $Z/Z_{\odot} = 0.1$, 
$1/3$ and $1$, respectively. Squares and circles represent the sdB stars analysed 
by Maxted et al. (\cite{max01}) and Lisker et al. (\cite{lisk05}), 
respectively. Filled symbols represent the sdB's with 
peculiar $\rm H_{\alpha}$ line profiles, which may indicate the presence of 
a weak wind.}
\end{figure}

In the figure it can be seen that the majority of sdB stars populate a region 
in the $T_{\rm eff}$ - $\log g$ diagram that is just below the line above 
which for $Z/Z_{\odot} = 1$ coupled winds may exist. If multicomponent 
effects are neglected, the results predict weak winds with mass-loss rates 
$\dot M \la 10^{-12} M_{\odot} / \rm yr$.
However, for such weak winds the momentum exchange between the metals, on 
the one hand, and hydrogen and helium, on the other, is not effective 
enough, because the densities in the wind are too low. 
Possible winds should be 
selective winds that lead to additional changes in the surface 
composition, which have not yet been taken into account in the diffusion 
calculations with and without mass-loss. 

According to the present results coupled winds can exist for the most luminous 
sdB stars, if the metallicity is not too low. This would explain why only 
in these stars wind signatures have been detected. 
The predicted mass-loss rates are of the order of $10^{-11}$ to 
$10^{-10} M_{\odot} / \rm yr$. This agrees with the predictions 
of the mass-loss recipe of Vink \& Cassisi (\cite{vink02}),
nevertheless the existence of these winds is still uncertain from the 
theoretical point of view. In Vink \& Cassisi's calculations, the 
terminal velocity has been a free parameter, and their results are 
for $v_{\infty} = v_{\rm esc}$. The assumption of higher terminal velocities 
would lead to lower mass-loss rates. In the present calculations, the 
finite disk correction and the shadowing of the flux by the photospheric 
lines have not been taken into account. Especially the latter effect 
seems to be very important. As discussed in Sect. 6, 
both simplifications should lead to the mass-loss rates 
being overestimated and the terminal velocities being underestimated. Then for 
a star with given stellar parameters and chemical composition, the  
densities in the wind would be lower. Consequently decoupling of the 
metals from hydrogen and helium would occur at lower gravities than 
predicted, and the lines shown in Fig. 13 
would be shifted to lower gravities. Frictional heating may have 
a similar effect. In addition, the presence of density inhomogenities in the 
wind, which are not taken into account in the present calculations, may possibly 
change the mass-loss rates. This effect of ``clumping" is still 
investigated (Oskinova et al. \cite{osk07}).
For luminous stars there is some evidence that the mass-loss rates  
may need to be revised downwards (e.g. Bouret et al. \cite{bou05}; 
Fullerton et al. \cite{ful06}; Martins et al. \cite{mar05}; 
Puls et al. \cite{puls06}).      

To confirm the existence of coupled winds in the 
most luminous sdB stars, more sophisticated calculations are necessary.
Because the sdB stars have a variety of abundances, it is important 
to show which elements contribute primarily to the radiative force. 
As explained by Puls et al. (\cite{puls00}), this should be the most 
abundant elements with the strongest lines in weak winds.
According to the present results, the most important element is carbon 
for $T_{\rm eff} = 25000$ and $30000 \, \rm K$. For higher 
effective temperatures the situation is less clear. 

If during the post-EHB evolution indeed winds with mass-loss rates  
$\dot M \ga 10^{-11} M_{\odot} / \rm yr$ were initiated, the 
outer layers of the star will be removed rapidly. This should lead to 
the abundance anomalies gradually decreasing, 
so elements that are not detectable in sdB stars within the EHB band 
may reappear on the surface during the post-EHB evolution. 
This may explain the result of 
Edelmann et al. (\cite{edel06}) that, in those sdB stars with 
$T_{\rm eff} \ga 32000 \, \rm K$ near the EHB band, no Si, Mg, and Al 
have been detected, whereas these elements are present in their more luminous 
counterparts with similar effective temperatures, which are in a post-EHB 
stage of evolution.

The results confirm the present picture of white-dwarf chemical evolution. 
The variety of compositions from hydrogen-rich to helium-carbon-oxygen-rich 
observed in pre-white dwarfs is due to the dredge up of processed material 
(Werner \& Herwig \cite{wer06}). Mass-loss rates between about $10^{-11}$ and 
$10^{-6} M_{\odot} / \rm yr$ are predicted for pre-white dwarfs with 
$M_{*} = 0.5 M_{\odot}$,  
$30000 \, \rm K \leq T_{\rm eff} \leq 50000 \, \rm K$, and $3.4 \la \log g \la 5.0$, 
dependent on stellar parameters and metallicities.   
These  mass-loss rates are large enough to prevent diffusion, at 
least if the metallicity is not reduced by more than about a factor of ten 
in comparison to the solar value. No wind solution exists for white dwarfs 
on the cooling sequence with similar 
effective temperatures. Even for solar metallicity, 
the maximum possible radiative acceleration is too low by about one order 
of magnitude. Thus somewhere during the evolution with 
$T_{\rm eff} > 50000 \, \rm K$, the mass-loss rates must decrease to very low 
values so that chemically homogeneous winds 
can no longer exist. The possibility that metallic winds still exist 
may explain the result of e.g. Good et al. (\cite{good05}) that 
neither diffusion calculations 
that assume chemically homogeneous winds nor diffusion calculations that 
assume the absence of any mass-loss can explain the measured abundances 
in DA and DAO white dwarfs very well.  

As discussed in Sect. 6.3, the existence of pure metallic winds with 
mass-loss rates $\dot M \la 10^{-16} M_{\odot} / \rm yr$ and with 
hydrostatic hydrogen and helium may be a promising scenario to explain 
the abundance anomalies of the metals in the various types of chemically 
peculiar stars considered in this paper. Then, however, 
the problem remains, why the abundances of helium in general are larger than 
predicted from diffusion calculations, which assume an equilibrium 
between gravitational settling and radiative levitation. Possibly 
the stellar atmospheres are not completely undisturbed after all.  

Hot (pre-) white dwarfs already have a variety of helium (and metal-) 
abundances before the onset of gravitational settling  
is expected. 
A dependence on the star's history may also play some role in 
sdB stars.
According to the analyses of Edelmann et al. (\cite{edel03}) and 
Lisker et al. (\cite{lisk05}) two distinct sequences of sdB stars seem to 
exist, which are characterised by an offset in the helium abundances. This 
phenomenon can hardly be explained with a single atmospheric effect.   
%%---------------------------------------------------------------------------------------------------------
%+++++++++++++++++++++++++++++++++++++++++++++++++++++++++++++++++++++++++++++++++++++
\begin{acknowledgements}
I would like to thank J. Vink for doing some additional mass-loss calculations, 
U. Heber for the calculation of a synthetic spectrum, A. Feldmeier for useful 
hints concerning the CAK theory, and I. Bues for carefully reading the manuscript.
The referee's comments have been a great help in improving the paper significantly.
\end{acknowledgements}

\end{document}